\documentclass[12pt]{article}

\usepackage{amsmath,amssymb,amsthm}
\usepackage{epsfig}
\usepackage{arydshln}

\newcommand{\bfm}[1]{\mbox{\boldmath ${#1}$}}
\newcommand{\nonum}{\nonumber \\}
\newcommand\eq[1] {(\ref{#1})} 
\newcommand{\beqa}{\begin{eqnarray}}
\newcommand{\eeqa}[1]{\label{#1}\end{eqnarray}}
\newcommand{\beq}{\begin{equation}}
\newcommand{\eeq}[1]{\label{#1}\end{equation}}
\newcommand{\Grad}{\nabla}
\newcommand{\Div}{\nabla \cdot}

\newcommand{\Real}{\mathop{\rm Re}\nolimits}
    \newcommand{\Imag}{\mathop{\rm Im}\nolimits}
\newcommand{\Tr}{\mathop{\rm Tr}\nolimits}

\newcommand{\lang}{\langle}
\newcommand{\rang}{\rangle}

\newcommand{\wt}{\widetilde}

\newcommand{\Ga}{\alpha}
\newcommand{\Gb}{\beta}
\newcommand{\Gd}{\delta}

\newcommand{\Gf}{\phi}

\newcommand{\Gc}{\chi}

\newcommand{\Gl}{\lambda}

\newcommand{\Gt}{\theta}

\newcommand{\Gr}{\rho}

\newcommand{\Gs}{\sigma}

\newcommand{\Gy}{\psi}

\newcommand{\GP}{\Pi}

\newcommand{\BGb}{\bfm\beta}

\newcommand{\BGs}{\bfm\sigma}

\newcommand{\BGG}{\bfm\Gamma}
\newcommand{\BGL}{\bfm\Lambda}
\newcommand{\BGP}{\bfm\Pi}

\newcommand{\BGU}{\bfm\Upsilon}

\newcommand{\BGX}{\bfm\Xi}
\newcommand{\BGY}{\bfm\Psi}

\newcommand{\CA}{{\cal A}}

\newcommand{\CD}{{\cal D}}
\newcommand{\CE}{{\cal E}}

\newcommand{\CH}{{\cal H}}

\newcommand{\CJ}{{\cal J}}

\newcommand{\CP}{{\cal P}}

\newcommand{\CR}{{\cal R}}
\newcommand{\CS}{{\cal S}}

\newcommand{\CU}{{\cal U}}

\newcommand{\bpm}{\begin{pmatrix}}
\newcommand{\epm}{\end{pmatrix}}
\newcommand\fig[1] {{\rm Figure}~\ref{fig:#1}}

\newcommand\labfig[1] {\label{fig:#1}}

\usepackage{url}
\usepackage{hyperref}

\def\Ba{{\bf a}}
\def\Bb{{\bf b}}

\def\Be{{\bf e}}

\def\Bg{{\bf g}}
\def\Bh{{\bf h}}

\def\Bj{{\bf j}}

\def\Bn{{\bf n}}

\def\Br{{\bf r}}

\def\Bu{{\bf u}}
\def\Bv{{\bf v}}
\def\Bw{{\bf w}}
\def\Bx{{\bf x}}

\def\BA{{\bf A}}
\def\BB{{\bf B}}

\def\BE{{\bf E}}

\def\BH{{\bf H}}
\def\BI{{\bf I}}
\def\BJ{{\bf J}}
\def\BK{{\bf K}}
\def\BL{{\bf L}}
\def\BM{{\bf M}}

\def\BP{{\bf P}}
\def\BQ{{\bf Q}}
\def\BR{{\bf R}}
\def\BS{{\bf S}}

\def\BU{{\bf U}}
\def\BV{{\bf V}}
\def\BW{{\bf W}}

\def\BY{{\bf Y}}
\def\BZ{{\bf Z}}

\newcommand{\rank}{\mbox{rank}}

\usepackage{graphics}
\usepackage{amssymb}

\title{Approximating the effective tensor as a function of the component
tensors in two-dimensional composites of two anisotropic phases}
\author{Graeme W. Milton}
\date{\small{Department of Mathematics, University of Utah, Salt Lake City, UT 84112, USA
\\Email: milton@math.utah.edu}}
\begin{document}
\maketitle
\vspace{2ex}
\begin{abstract}
A conducting two-dimensional periodic composite of two anisotropic phases with anisotropic,
not necessarily symmetric, conductivity tensors $\BGs_1$ and $\BGs_2$ is considered.
By finding approximate representations for the relevant operators, an approximation
formula is derived for the effective matrix-valued conductivity $\BGs^*$ as a function
of the two matrix-valued conductivity tensors $\BGs_1$ and $\BGs_2$. This approximation
should converge to the exact function $\BGs^*(\BGs_1,\BGs_2)$ as the number of basis fields 
tends to infinity. Using the approximations for the relevant operators one can also directly
obtain approximations, with the same geometry, for the effective tensors $\BL^*$ of coupled field problems, including elasticity,
piezoelectricity, and thermoelectricity. To avoid technical complications we assume that the phase geometry is symmetric 
under reflection about one of the centerlines of the unit cell. 
\end{abstract}
\vspace{3ex}
\section{Introduction}
\setcounter{equation}{0}

In a periodic composite of two anisotropic phases the conductivity tensor takes the form
\beq \BGs(\Bx)=\Gc(\Bx)\BGs_1+[1-\Gc(\Bx)]\BGs_2, \eeq{0.0}
where the indicator function $\Gc(\Bx)$ is periodic and 1 in phase 1 and 0 in phase 2,
and $\BGs_1$ and $\BGs_2$ are the matrix-valued conductivity tensors of the two phases:
they are $2\times 2$ matrices for two-dimensional composites, and $3\times 3$ matrices for three-dimensional composites. The effective
conductivity tensor is found by looking for current fields $\Bj(\Bx)$ and electric fields $\Be(\Bx)$, with the same periodicity of the composite,
that solve
\beq \Bj(\Bx)=\BGs(\Bx)\Be(\Bx),\quad \Div\Bj=0,\quad \Be=-\Grad V(\Bx). \eeq{0.0aa}
In these equations $V(\Bx)$ is the electric potential, and the volume average, $\lang\Be\rang$, of the electric field $\Be(\Bx)$ is prescribed. Here and
later the angular brackets $\lang\cdot\rang$ denote an average over the unit cell. The average current field $\lang\Bj\rang$ depends linearly
on $\lang\Be\rang$, and it is this linear relation,
\beq \lang\Bj\rang=\BGs^*\lang\Be\rang, \eeq{0.0ab}
that determines the effective tensor $\BGs^*$. It clearly depends on $\BGs_1$ and $\BGs_2$, as well as the geometry $\Gc(\Bx)$ of the
composite which we assume to be fixed.
The effective conductivity $\BGs^*(\BGs_1,\BGs_2)$ is then, respectively, a $2\times 2$ or $3\times 3$ matrix-valued
analytic function of $\BGs_1$ and $\BGs_2$, in the sense of being an analytic function of all the elements 
of $\BGs_1$ and of all the elements of $\BGs_2$, in the domain where the $\Real(\BGs_1)>0$ and $\Real(\BGs_2)>0$.
It is also a homogeneous function in the sense that 
\beq \BGs^*(\Gl\BGs_1,\Gl\BGs_2)=\Gl\BGs^*(\BGs_1,\BGs_2), \eeq{0.0a}
for all real or complex $\Gl$. Consequently, the region of analyticity can be extended to the union over $\Gt$
of the domains where $\Real(e^{i\Gt}\BGs_1)>0$ and $\Real(e^{i\Gt}\BGs_2)>0$.

Here we address the question: How can one approximate the function $\BGs^*(\BGs_1,\BGs_2)$? One approach
that has proved very successful in many different contexts is to use an effective medium approximation. In the case
of same shaped inclusions of phase 1 dispersed in a matrix of phase 2 one may, following the approach
of \cite{Frohlich:1946:TRP}, take an inclusion of phase 1,
and surround it by a coating of phase 2 (with thickness chosen so the relative volume fractions of the phases in the
coated inclusions matches that in the original composite) and embed this coated inclusion in an infinite medium 
of conductivity $\BGs^*$ in which a uniform field at infinity is applied. The value of $\BGs^*$ is chosen so the
leading perturbation to the field at large distances (the dipolar field) vanishes for all directions of the applied field.
This is a realizable approximation formula in the sense that there corresponds a hierarchical 
geometry whose conductivity function $\BGs^*(\BGs_1,\BGs_2)$ corresponds exactly to that of the
approximation \cite{Milton:1984:CEE,Milton:1985:TCP,Avellaneda:1987:IHD}. However, the solution to the
problem of a coated inclusion in an infinite medium is not necessarily easy, and may have to be done numerically.
Originally developed for composites with isotropic constituents \cite{Bruggeman:1935:BVP}, the effective medium approximation
has been extended to the case in which the constituents are anisotropic \cite{Stroud:1975:GEM,Helsing:1991:ECA,Levy:2013:EMA}, and has been 
very successfully applied to dynamic problems well beyond the quasistatic limit as well \cite{Wu:2007:CSC,Xhang:2015:EMT}, 
although in the dynamic case it seems unlikely that it is realizable.

There are also generalizations of the coated sphere and coated disk assemblage model of Hashin \cite{Hashin:1962:EMH}
comprised of assemblages of coated ellipsoids or ellipses of an anisotropic material, coated by an isotropic phase, with the inner and
outer surfaces being confocal, that are then stretched to achieve any desired real or complex value of the conductivity matrix $\BGs_1$ of
the inclusion phase and any real positive definite tensor $\BGs_2$ of the matrix phase. While one can obtain an explicit and exact formula for
the effective tensor $\BGs^*$ of these assemblages [see, for example, formula (5.95) in \cite{Sihvola:1999:EMF} and
section 8.4 of \cite{Milton:2002:TOC}] one does not get a function $\BGs^*(\BGs_1,\BGs_2)$ for a fixed geometry, as the assemblage geometry 
depends on $\BGs_2$. Moreover, $\BGs_2$ is not allowed to be complex in these geometries.

One can, of course, get an explicit formula for $\BGs^*(\BGs_1,\BGs_2)$ for simple laminate geometries, and associated hierarchical laminates that
are obtained by successive laminations on larger and larger length scales \cite{Maxwell:1954:TEM,Tartar:1979:ECH,Tartar:1985:EFC}. Numerical
methods have also been developed for solving for the effective tensor $\BGs^*$ with an anisotropic inclusion and an anisotropic matrix \cite{Helsing:1995:EAI}.

Our aim here, however, is to develop approximations that have the capability of fitting the exact function 
$\BGs^*(\BGs_1,\BGs_2)$ arbitrarily closely for a wide variety of composite geometries.
The approach we take is to find canonical approximate representations
of all the relevant operators as finite-dimensional matrices and then use these to develop an approximation
for the function $\BGs^*(\BGs_1,\BGs_2)$. A similar approach was developed for the effective conductivity
of composites of $n$ isotropic phases, having scalar conductivities $\Gs_1,\Gs_2,\ldots,\Gs_n$, and it lead to 
a new type of continued fraction expansion for the function $\BGs^*(\Gs_1,\Gs_2,\ldots,\Gs_n)$
(involving matrices of increasing dimension as one proceeded down the continued fraction) that in turn
produced a variety of new bounds on the effective conductivity tensor $\BGs^*$ \cite{Milton:1987:MCEa,Milton:1987:MCEb}. 
This method for deriving bounds was called the field equation recursion method: see Chapter 29 in \cite{Milton:2002:TOC}
and Chapters 9 and 10 in \cite{Milton:2016:ETC}. It was observed (see Appendix E in \cite{Milton:1981:BTO}, \cite{Kantor:1984:IRB}, \cite{DellAntonio:1986:ATO}, 
Section 15 in \cite{Milton:1987:MCEa}, \cite{Barabash:1999:SRE}, and Section 18.2 and Chapter 29 in \cite{Milton:2002:TOC})
that the same approach allows one to easily obtain bounds on the possibly complex effective conductivity tensor or
effective elasticity tensor of multicomponent media
with anisotropic phases. For example, if we are in three-dimensions and there are two phases, possibly in many different orientations as in a two-phase polycrystal, then $\Gs_1$, $\Gs_2$, and $\Gs_3$ can be taken as the eigenvalues of the conductivity tensor of one phase,
while $\Gs_4$, $\Gs_5$, and $\Gs_6$ can be taken as the eigenvalues of the conductivity tensor of the second phase,
assuming in the case of complex conductivities that the real and imaginary parts of the conductivity tensor have the same
eigenvectors. Explicit bounds on the complex effective conductivity (or dielectric) tensor for multicomponent media with possibly 
anisotropic phases can also be obtained using the variational principles of Cherkaev and Gibiansky \cite{Cherkaev:1994:VPC}:
see formula (16.45) in \cite{Milton:1990:CSP}. Such bounds have application to determining the effective complex dielectric tensor
of sea ice as it can be regarded as a polycrystalline material \cite{Gully:2015:BCP}. In fact, the variational 
principles of Cherkaev and Gibiansky \cite{Cherkaev:1994:VPC}
also allow one to bound the viscoelastic moduli of composites \cite{Gibiansky:1993:EVM,Milton:1997:EVM,Gibiansky:1999:EVM}.

For simplicity, we focus on the two-dimensional case as the three-dimensional case would be exceedingly more challenging. Also, for simplicity, we assume that the phase geometry
as determined by the function $\Gc(\Bx)$ is symmetric under reflection about the vertical centerline of the unit
cell: if the origin is chosen so this is the line $x_1=0$ we require that $\Gc(-x_1,x_2)=\Gc(x_1,x_2)$. An
example of such a geometry is shown in \fig{1}. This extra symmetry allows us to avoid some technical complications.
\begin{figure}
	\centering
	\includegraphics[width=0.9\textwidth]{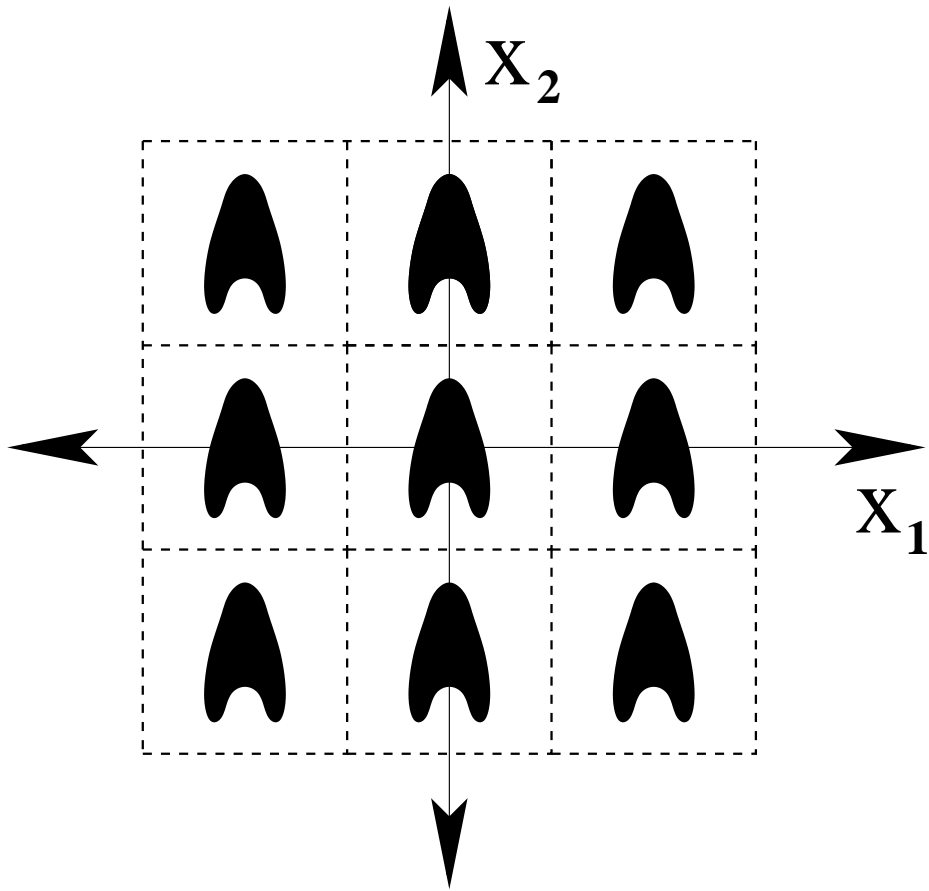}
	\caption{An example of a two-dimensional, two-phase geometry to which our analysis applies. Note that the geometry is symmetric about the vertical 
centerline of the unit cell. The periodic pattern should of course extend to infinity in both directions.}
	\labfig{1}
\end{figure}

The approach we take of finding appropriate representations for the relevant operators has the advantage that we can also directly find the associated
approximations for the effective tensors of other coupled field problems, such as elasticity or piezoelectricity problems. There the equations
take the form
\beq \BJ(\Bx)=\BL(\Bx)\BE(\Bx), \quad \BL(\Bx)=\Gc(\Bx)\BL_1+[1-\Gc(\Bx)]\BL_2,\quad \Div\BJ=0,\quad \BE=\Grad\BV, \eeq{0.0ac}
where $\BV$ is an $k$-component vector potential, $\BJ(\Bx)$ and $\BE(\Bx)$ are $d\times k$ matrix-valued fields with the same periodicity
as $\Gc(\Bx)$ (the divergence $\Div$ acts on the first index of $\BJ(\Bx)$) and $\BL(\Bx)$ is a 4-index object (a fourth-order tensor in the case of elasticity). Again it is the linear relation
\beq \lang\BJ\rang=\BL^*\lang\BE\rang, \eeq{0.0ad}
that determines the effective tensor $\BL^*$. The connection with conductivity is more apparent if we rewrite the equations in the form
\beq  \bpm \Bj^{(1)}(\Bx) \\ \Bj^{(2)}(\Bx) \\ \vdots \\ \Bj^{(k)}(\Bx)\epm=\bpm \BGs^{(11)}(\Bx) & \BGs^{(12)}(\Bx) & \ldots & \BGs^{(1k)}(\Bx) \\
\BGs^{(21)}(\Bx) & \BGs^{(22)}(\Bx) & \ldots & \BGs^{(2k)}(\Bx) \\
\vdots & \vdots & \ddots & \vdots \\
\BGs^{(k1)}(\Bx) & \BGs^{(k2)}(\Bx) & \ldots & \BGs^{(kk)}(\Bx) \epm
\bpm \Be^{(1)}(\Bx) \\ \Be^{(2)}(\Bx) \\ \vdots \\ \Be^{(k)}(\Bx)\epm
\eeq{0.0ae}
where the $\Bj^{(i)}(\Bx)$ and $\Be^{(j)}(\Bx)$ are the columns of $\BE(\Bx)$ and $\BJ(\Bx)$, respectively, 
\beq \Div\Bj^{(i)}=0,\quad \Be^{(j)}=\Grad V_j,\quad \BGs^{(ij)} (\Bx)=\Gc(\Bx)\BGs^{(ij)}_1+[1-\Gc(\Bx)]\BGs^{(ij)}_2,
\eeq{0.0af}
and
\beq \{\BGs^{(ij)}_1\}_{pq}=\{\BL_1\}_{piqj},\quad \{\BGs^{(ij)}_2\}_{pq}=\{\BL_2\}_{piqj}. \eeq{0.0ag}

Even in two-dimensions the function $\BGs^*(\BGs_1,\BGs_2)$ can be regarded as four scalar-valued functions
(the elements of $\BGs^*$) each depending on  eight variables (the matrix elements of $\BGs_1$ and $\BGs_2$). Using some
equivalences \cite{Dykhne:1970:APR} it is possible to transform any problem where $\BGs_1$ and $\BGs_2$ are not symmetric matrices
into an equivalent problem where both $\BGs_1$ and $\BGs_2$, and hence $\BGs^*$ are symmetric matrices \cite{Milton:1988:CHE}.
Thus the conductivity function $\BGs^*(\BGs_1,\BGs_2)$, where $\BGs_1$ and $\BGs_2$ are not necessarily symmetric, 
can be easily recovered from the function restricted to symmetric $\BGs_1$ and $\BGs_2$.
Also due to the homogeneity \eq{0.0a} we can assume, say, that one matrix element, such as $\{\BGs_1\}_{11}$, is unity.
We are still left with approximating three scalar-valued functions (the elements of a symmetric $\BGs^*$)
each depending on 5 complex variables (the matrix elements of symmetric matrices $\BGs_1$ and $\BGs_2$, excluding $\{\BGs_1\}_{11}$), which remains a
daunting task. It is this task that we will address.
\section{A review of some previous results}
\setcounter{equation}{0}
For composites of two phases having isotropic conductivities $\Gs_1$ and $\Gs_2$  the analytic properties of the effective
conductivity were first investigated by Bergman \cite{Bergman:1978:DCC}.  The formulation allowed Bergman to recover the famous Hashin-Shtrikman bounds on
the effective conductivity, dielectric constant, or magnetic permeability, of an isotropic composite of the two phases \cite{Hashin:1962:VAT}, 
as well as to obtain some new bounds.
There were some flaws in his arguments \cite{Milton:1979:TST} that could be corrected by 
approximating the composite by a periodic network of two impedances for which the effective conductivity is a rational function of $\Gs_1$ and $\Gs_2$ 
\cite{Milton:1981:BCP}. Any diagonal element $\Gs^*_{ii}$ (no sum over $i$) of the
effective conductivity tensor $\BGs^*$ expressed as a function of the component conductivities has the approximation formula
\beq \Gs^*_{ii}(\Gs_1,\Gs_2)=\sum_{\Ga=0}^{m+1}\frac{a_{\Ga}}{q_\Ga/\Gs_1+(1-q_\Ga)/\Gs_2}, \eeq{0.1}
with parameters satisfying the constraints
\beq 1=q_0>q_1>q_2>\ldots>q_m>q_{m+1}=0,\quad a_\Ga\geq 0~~{\rm for~all~}\Ga,\quad\sum_{\Ga=0}^{m+1}a_{\Ga}=1. \eeq{0.2}
This representation, for rational functions $\Gs^*_{ii}(\Gs_1,\Gs_2)$, \eq{0.1} is an immediate consequence of the homogeneity property
\eq{0.0a}, the Nevanlinna-Herglotz type property, 
\beq \Real{\BGs^*(\Gs_1,\Gs_2)}\geq 0~ {\rm when}~ \Real\Gs_1\geq 0~ {\rm and}~ \Real\Gs_1\geq 0,
\eeq{0.2a}
and the normalization property that $\BGs^*(1,1)=\BI$. The key observation that leads to the representation is that around pole of order $m$ of
any rational function, and in particular $\Gs^*_{ii}(\Gs_1,i)=-\Gs^*_{ii}(-\Gs_1,-i)$, its real part changes sign $m$ times. Note that one of either $\Gs_1$ or $-\Gs_1$ has non-negative real part so we can apply \eq{0.2a}. Therefore 
the poles of $\Gs^*_{ii}(\Gs_1,i)$ must be simple and lie on the imaginary $\Gs_1$-axis, and not on the positive imaginary axis
as then $\Gs^*_{ii}(\Gs_1,1)=-i\Gs^*_{ii}(i\Gs_1,i)$ would violate \eq{0.2a} in the vicinity of these poles.
The representation also follows from standard representation formulas of Nevanlinna-Herglotz and Stieltjes
functions applied to the rational Nevanlinna-Herglotz function $f(z)=\Gs^*_{ii}(z,1)$ with $z=\Gs_1/\Gs_2$ or, more specifically, to the rational 
Stieltjes function $g(z)=\Gs^*_{ii}(1/z,1)$ with $z=\Gs_2/\Gs_1$. For such rational functions, the corresponding measures entering their integral 
representation formulas are discrete and one easily obtains \eq{0.1}.

In fact, it is clear that there is always a geometry, shown in \fig{2}, that has the function $\Gs^*_{ii}(\Gs_1,\Gs_2)$ given by \eq{0.1} as a diagonal element of its effective
conductivity function: as shown in Appendix B of \cite{Milton:1981:BTO}, one may first obtain $m+2$ laminates of the two phases mixed in proportions $q_\Ga$ and $1-q_\Ga$
(including the pure phases as $q_0=1$ and $q_{m+1}=0$) and then one can slice together these laminates in the orthogonal direction in
proportions $a_{\Ga}$, $\Ga=0,1,2,\ldots,m+1$. 
\begin{figure}
	\centering
	\includegraphics[width=0.9\textwidth]{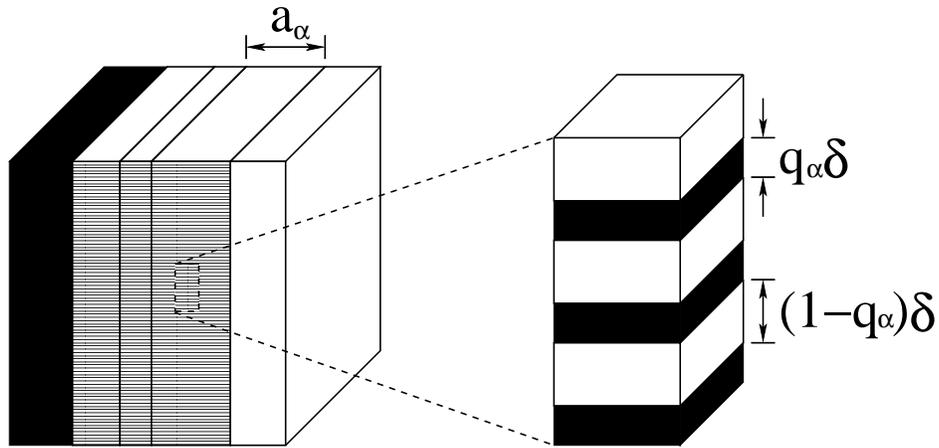}
	\caption{An example of a two-scale laminate built from two phases, phase 1 in black and phase 2 in white, 
that has an effective conductivity in the direction of the $x_1$ axis given by the formula \protect{\eq{0.1}} in the limit $\Gd\to 0$. 
Only one unit cell is shown: the structure is repeated periodically.}
	\labfig{2}
\end{figure}

Bounds were obtained on the complex conductivity $\BGs^*$ for complex values of $\Gs_1$ and $\Gs_2$, or equivalently
on the complex dielectric constant for complex values of the component dielectric constants 
\cite{Milton:1979:TST,Milton:1980:BCD,Milton:1981:BCP,Bergman:1980:ESM,Bergman:1982:RBC}. These also
lead to bounds on the complex polarizability of an isolated inclusion (see figure 3 in \cite{Milton:1981:TPA}, and \cite{Milton:2017:BCP}) that have
seen a resurgence of interest due to relevance in determining the maximum screening properties of a smoke screen of
small metal particles \cite{Miller:2014:FLE} (see also \cite{Miller:2015:FLO} for bounds that do not assume a quasistatic approximation).
In fact a whole hierarchy of bounds were obtained on the diagonal element $\Gs^*_{ii}$ for specific
values of $\Gs_1$ and $\Gs_2$ given partial information about the function $\BGs^*(\Gs_1,\Gs_2)$ \cite{Milton:1981:BTO}.
In a wider context, many of these bounds correspond to known bounds on Stieltjes functions: see \cite{Milton:1986:MPC}, 
the discussion in the introduction of \cite{Milton:1987:MCEb} and references therein, and Chapter V in
\cite{Krein:1977:MMP}. As kindly pointed out to me by Mihai Putinar, there is also close connection with Nevanlinna-Pick interpolation
\cite{Nevanlinna:1919:BFG,Nevanlinna:1929:BAF,Pick:1916:UBA}.
Many of the bounds can also be obtained from bounds on variational principles: see \cite{Milton:1982:CTM}.

As $m\to\infty$ the approximations \eq{0.1} converge to an integral representation formula for the effective 
conductivity diagonal element $\Gs^*_{ii}(\Gs_1,\Gs_2)$ that has been rigorously established by Golden and Papanicolaou \cite{Golden:1983:BEP}.

In three dimensions, even with only two isotropic components, the complete set of constraints on the conductivity function is still unknown. For isotropic composites of two
isotropic phases having conductivities $\Gs_1$ and $\Gs_2$ the effective conductivity tensor takes the form $\BGs^*(\Gs_1,\Gs_2)=\Gs^*(\Gs_1,\Gs_2)\BI$
and it has been established \cite{Milton:1981:BCP,Avellaneda:1988:ECP,Nesi:1991:MII,Zhikov:1991:EHM,Zhikov:1992:EEH} that the scalar conductivity 
function $\Gs^*(\Gs_1,\Gs_2)$ satisfies the phase interchange inequality
\beq \frac{\Gs^*(\Gs_1,\Gs_2)\Gs^*(\Gs_2,\Gs_1)}{\Gs_1\Gs_2}+\frac{\Gs^*(\Gs_1,\Gs_2)+\Gs^*(\Gs_2,\Gs_1)}{\Gs_1+\Gs_2} \geq 2.
\eeq{0.7}
It seems likely that there could be other, as yet undiscovered, constraints.
When the composite is possibly anisotropic, analogously to \eq{0.1}, one has an approximation for the matrix-valued effective conductivity
\beq \BGs^*(\Gs_1,\Gs_2)=\sum_{\Ga=0}^{m+1}\frac{\BA_{\Ga}}{q_\Ga/\Gs_1+(1-q_\Ga)/\Gs_2}, \eeq{0.7a}
where the parameters $q_0$, $q_1$ ,$\ldots$, $q_{m+1}$ and symmetric three-by-three matrices $\BA_i$ satisfy the constraints 
\beq 1=q_0>q_1>q_2>\ldots>q_m>q_{m+1}=0,\quad \BA_\Ga\geq 0~~{\rm for~all~}\Ga,\quad\sum_{\Ga=0}^{m+1}\BA_{\Ga}=1. \eeq{0.7b}
Again, this is an immediate consequence of the homogeneity property \eq{0.0a}, the Nevanlinna-Herglotz type property, \eq{0.2a}
and the normalization property that $\BGs^*(1,1)=\BI$ (and, conversely, any function of the form \eq{0.7a} such
that \eq{0.7b} holds, necessarily satisfies these three properties).
When $\Gd\Gs=\Gs_1-\Gs_2$ is small, the first terms in the expansion of $\BGs^*(\Gs_2+\Gd\Gs,\Gs_2)$, given by \eq{0.7a} in powers of $\Gd\Gs$ are
\beq \BGs^*(\Gs_2+\Gd\Gs,\Gs_2)=\Gs_2\BI+\Gd\Gs\sum_{\Ga=0}^{m+1}q_\Ga\BA_{\Ga}-(\Gd\Gs)^2\sum_{\Ga=0}^{m+1}q_\Ga(1-q_\Ga)\BA_{\Ga}/\Gs_2+\ldots,
\eeq{0.7c}
and we have the sum rules \cite{Brown:1955:SMP,Bergman:1978:DCC,Kohn:1986:BEC},
\beq \sum_{\Ga=0}^{m+1}q_\Ga\BA_{\Ga}=f\BI,\quad \sum_{\Ga=0}^{m+1}q_\Ga(1-q_\Ga)\Tr[\BA_{\Ga}]=f(1-f),
\eeq{0.7d}
where $f$ can be identified with the volume fraction of phase $1$. Even if $f$ is unknown, the sum rules clearly constrain the possible values of the set of 
$q_\Ga$ and $\BA_\Ga$, $\Ga=0,1,2,\ldots,m+1$. There are additional constraints. For example, given the material with anisotropic effective conductivity
$\BGs^*(\Gs_1,\Gs_2)$, a result of Schulgasser \cite{Schulgasser:1977:BCS}
shows that there is an isotropic composite, constructed from the material with conductivity $\BGs^*(\Gs_1,\Gs_2)$,
that has a scalar effective conductivity
\beq \Gs^*(\Gs_1,\Gs_2)=\Tr[\BGs^*(\Gs_1,\Gs_2)]/3, \eeq{0.7e}
and this must satisfy \eq{0.7}.

For two-dimensional anisotropic composites of two isotropic phases with conductivities $\Gs_1$ and $\Gs_2$ one can completely
characterize the possible effective conductivity functions $\BGs^*(\Gs_1,\Gs_2)$ \cite{Milton:1986:APLG} (see also Section 18.5 in \cite{Milton:2002:TOC}.
The function $\BGs^*(\Gs_1,\Gs_2)$ is known to satisfy the Keller-Dykhne-Mendelson
phase interchange relationship \cite{Keller:1964:TCC,Dykhne:1970:CTD,Mendelson:1975:TEC}, 
\beq \BGs^*(\Gs_2,\Gs_1)=\Gs_1\Gs_2\BR_\perp[\BGs^*(\Gs_1,\Gs_2)]^{-1}\BR_\perp^T, \eeq{0.2aa}
where $\BR_\perp$, with transpose $\BR_\perp^T$ is the matrix for a $90^\circ$ rotation:
\beq
\BR_\perp= \bpm 0 & -1\\ 1 & 0\epm. \eeq{0.2b}
The approximation \eq{0.7a} for $\BGs^*(\Gs_1,\Gs_2)$ still holds, but now with symmetric two-by-two matrices $\BA_i$ satisfy the constraints
\eq{0.7b} and \eq{0.7d}. Let us exclude the trivial case where $\BGs^*(\Gs_1,\Gs_2)=\Gs_2\BI$ for all $\Gs_1$.
Then, to account for the phase interchange relationship, it is better to assume by homogeneity that $\Gs_2=1$, introduce
the variable $s=1/(1-\Gs_1)$ \cite{Bergman:1978:DCC}, and
obtain an approximation for
\beq \BS^*(s)=[\BI-\BGs^*(1-1/s,1)]^{-1}. \eeq{0.3}
The inverse in \eq{0.3} exists for all complex
$s$ not in the interval $[0,1]$ on the real axis. Indeed, with our exclusion, $f\ne 0$ and 
the approximation \eq{0.7a}, with \eq{0.7b} and \eq{0.7d}, implies that  $\BGs^*(\Gs_1,1)$ is a strictly
monotonic function of real $\Gs_1$ for $\Gs_1\in (0,\infty)$ (in the sense that the derivative is a strictly positive definite matrix) and
has a strictly positive definite imaginary part when $\Imag \Gs_1\ne 0$. These facts imply the invertibility of $[\BI-\BGs^*(1-1/s,1)]$
for $s$ not in the interval $[0,1]$ on the real axis. It is then straightforward to show $\BS^*(s)$ satisfies the analytic properties
\beqa &~&\BS^*(s)+\BR_\perp\BS^*(1-s)\BR_\perp^T=\BI, \quad\quad \BS^*(\overline{s})=\overline{\BS^*(s)},  \nonum
&~& \BS^*(s)> \BI~~~{\rm for~all~real~}s> 1, \quad\quad \BS^*(s)< 0~~~{\rm for~all~real~}s< 0, \nonum
&~&\Imag[\BS^*(s)]> 0~~~{\rm whenever}~~\Imag(s)> 0,
\eeqa{0.4}
the first of which is implied by \eq{0.2aa}.
Any rational approximation to $\BS^*(s)$ satisfying these properties takes the form
\beq \BS^*(s)\approx s(1+\Tr\BA)\BI-\BA+\sum_{i=1}^m\frac{\BS_i}{s_i-s}+\frac{\BR_\perp^T\BS_i\BR_\perp}{(1-s_i)-s}, \eeq{0.5}
where the poles $s_i$, the $2\times 2$ symmetric residue matrices $\BS_i$, and the $2\times 2$ symmetric matrix $\BA$ 
can all be freely chosen subject to the restrictions that 
\beq 1>s_i>0, \quad \BS_i\geq 0,\quad \BA\geq \sum_{i=1}^m\frac{\BS_i}{s_i}+\frac{\BR_\perp^T\BS_i\BR_\perp}{1-s_i}. \eeq{0.6}
This is easy to establish: the first equation in \eq{0.4} shows that a singularity at $s=s_i$ must be accompanied by
a singularity at $s=1-s_i$ and that the matrix-valued residues must be related as in \eq{0.5}; similarly, to satisfy \eq{0.4}, the
constant term and the term proportional to $s$ in the expression for $\BS^*(s)$ must be related as in the expansion \eq{0.5}; and,
finally, the restriction that $\BS^*(s)$ remains negative definite as $s\to 0$ from below (implied by the fact that $\BGs^*(\Gs_1,1)\geq\BI$
when $\Gs_1>1$) gives the last restriction in \eq{0.6}. 
There is a one-to-one correspondence between rational functions satisfying the constraints \eq{0.4}
and hierarchical laminate geometries, illustrated in \fig{3}, obtained by starting with a core material of either phase 1 or phase 2,
and successively laminating it, in varying directions, on larger and larger length scales, with alternating 
choices of phase 1 and phase 2 \cite{Milton:1986:APLG}. This and related representation formulae have lead
to a variety of sharp bounds on the effective conductivity tensor $\BGs^*$
of two-dimensional composites of two isotropic phases \cite{Milton:1980:BCD,Milton:1981:BCP,Clark:1995:OBC}. Some of the bounds
had been obtained first from variational principles \cite{Cherkaev:1992:ECB}.  

\begin{figure}
	\centering
	\includegraphics[width=0.9\textwidth]{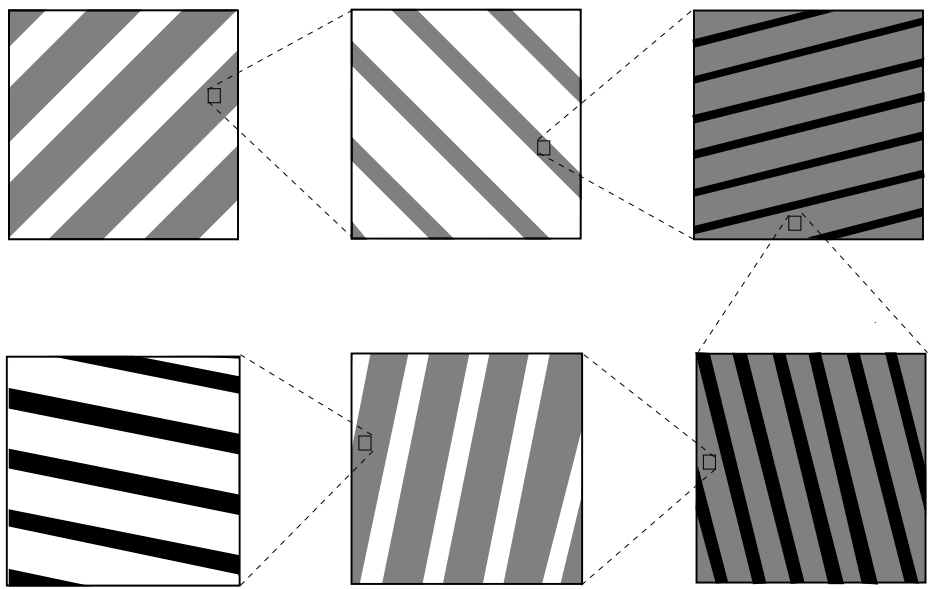}
	\caption{An example of a hierarchical laminate built from two isotropic phases, with phase 1 in black and phase 2 in white. The gray
regions represent the hierarchical laminate at the previous step. In two-dimensions, these form a representative class of geometries that can mimic any matrix-valued
effective conductivity function $\BGs^*(\Gs_1,\Gs_2)$ of the conductivities $\Gs_1$ and $\Gs_2$ of the two phases.}
	\labfig{3}
\end{figure}

For a polycrystal, namely a composite of a single crystal in many different orientations, the local conductivity takes the form
\beq \BGs(\Bx)=\BR^T(\Bx)\BGs_0\BR(\Bx), \eeq{0.7f}
in which $\BR(\Bx)$ is a rotation field, satisfying $\BR^T(\Bx)\BR(\Bx)=\BI$, that is also periodic in $\Bx$. A natural problem is then how to approximate 
the function $\BGs^*(\BGs_0)$ which is a two-by-two matrix-valued function of four variables, these being the matrix values of $\BGs_0$. A corollary of the
work of Clark and Milton \cite{Clark:1994:MEC,Clark:1997:CFR} is that this function can be approximated, arbitrarily closely, by the function associated with a polycrystal which is a hierarchical sequential laminate, as illustrated in \fig{4}

\begin{figure}
	\centering
	\includegraphics[width=0.9\textwidth]{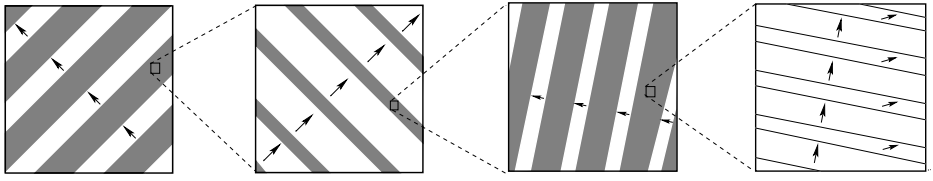}
	\caption{An example of a hierarchical laminate built from a single crystal placed in different orientations. With respect to some basis the crystal orientation
is denoted by the arrow. The gray regions represent the hierarchical laminate at the previous step. 
In two-dimensions, these form a representative class of geometries that can mimic any matrix-valued
effective conductivity function $\BGs^*(\BGs)$ of the conductivity matrix $\BGs_0$ of the single crystal.}
	\labfig{4}
\end{figure}

The $n$-th order approximation for the effective conductivity tensor $\BGs_n^*$ then involves an arbitrary unit vector $\Bn_0$ that we are free to choose, an associated
series of $n+1$ two-by-two rotation matrices $\BR_0$, $\BR_1$, $\ldots$, $\BR_n$
each satisfying $(\BR_j)^T\BR_j=\BI$, where $\BA^T$ denotes the transpose of the matrix $\BA$,
and an associated sequence of $n$ volume fractions $p_1$, $p_2$, $\ldots$, $p_n$, satisfying $0 <p_j<1$, for $j=1,2,\ldots,n$. 
The expression for $\BGs^*_n(\BGs_0)$ as a function of $\BGs_0$ is provided, for example, by iterating the lamination formula (9.18) in
\cite{Milton:2002:TOC} (derived in \cite{Milton:1990:CSP} and \cite{Zhikov:1991:EHM}). The result is easily expressed in terms of a 
matrix-valued function $\BS^*_n(\BS)$, where 
\beq \BS=\Gs^0[\Gs^0\BI-\BGs_0]^{-1}, \eeq{0.8}
and $\Gs^0$ is an arbitrary constant, through the formula
\beq \BGs^*_n(\BGs_0)=\Gs^0\BI-\Gs^0[\BS^*_n(\BS)]^{-1}, \eeq{0.9}
where $\BS^*_n(\BS)$ is obtained iteratively through the equations
\beq [\BS^*_j-\Bn_j(\Bn_j)^T]^{-1}=p_j[(\BR_j)^T\BS\BR_j-\Bn_j(\Bn_j)^T]+(1-p_j)[\BS^*_{j-1}-\Bn_j(\Bn_j)^T]^{-1},
\eeq{0.10}
for $j=n,n-1,\ldots, 1$ with 
\beq \Bn_j=(\BR_j)^T\Bn_0~~{\rm for}~j\geq 1,\quad \BS^*_0=(\BR_0)^T\BS\BR_0. \eeq{0.11}
Even though we have an essentially complete characterization of what conductivity functions 
$\BGs^*(\BGs_0)$ are possible for two-dimensional polycrystals, it is still not known if there is a 
simple integral representation formula for this function.

For broader surveys of the analytic properties of the effective moduli of composites and
on bounds on these effective moduli the reader is referred to the books 
\cite{Cherkaev:2000:VMS,Torquato:2001:RHM,Allaire:2000:SOH,Milton:2002:TOC,Tartar:2009:GTH,Milton:2016:ETC}.

To avoid excessive notation we will typically use the same symbol for an operator and for the matrix that represents it in a basis.
In the rare case that we change basis, we will spell out how the matrix representing the operator changes. Also, $\BA^T$ 
will represent either the adjoint of the operator $\BA$, or equivalently the transpose of the matrix representing $\BA$  with
respect to an orthonormal basis, i.e., that matrix representing the adjoint operator.
\section{Conductivity of two-phase media, one isotropic the other anisotropic}
\setcounter{equation}{0}
We consider a two-dimensional problem of, e.g., anisotropic inclusions in
an isotropic matrix. We assume that the geometry has reflection
symmetry, i.e., is invariant under the transformation $x_1\to - x_1$. 
We also begin by 
assuming that the axis of the anisotropic phase is aligned with the
coordinate axis (this assumption will be removed in the Section 4 of the paper).
Thus, we have
\beq
\BGs(\Bx) =\BGs_1\Gc(\Bx)+\BGs_2(1-\Gc(\Bx))\text{ with }
\BGs_1=\bpm
\Gl_1 & 0\\ 0 & \Gl_2
\epm,\quad
\BGs_2
\bpm
\Gl_3 & 0 \\ 0 & \Gl_3
\epm,
\eeq{1}
where
\beq
\Gc(\Bx) = \left\{\begin{array}{l}
1 \quad\text{in phase 1 (the inclusions),}\\
0 \quad\text{in phase 2 (the matrix).}
\end{array}
\right.
\eeq{2}
The assumed reflection symmetry of the geometry implies the associated effective tensor is diagonal:
\beq  \BGs^* = 
\bpm
\Gs_{11}^* & 0\\ 0 & \Gs_{22}^*
\epm,
\eeq{2.a}
where the Mendelson duality relationship \cite{Mendelson:1975:TEC} implies
that when these diagonal elements are considered as functions of $\Gl_1$, $\Gl_2$, and $\Gl_3$, they satisfy
\beq  \Gs_{22}^*(\Gl_1,\Gl_2,\Gl_3)=\frac{1}{\Gs_{11}^*(1/\Gl_2,1/\Gl_1,1/\Gl_3)}.
\eeq{2.b}
Thus to determine $\BGs^*(\Gl_1,\Gl_2,\Gl_3)$ it suffices to know the function $\Gs_{11}^*(\Gl_1,\Gl_2,\Gl_3)$.

In this section we first introduce the relevant operators important to the analysis and list their algebraic properties.
Then in a truncated Hilbert space, with a carefully chosen basis, we obtain matrix representations for the relevant operators
and this then leads to an approximation for the function $\Gs_{11}^*(\Gl_1,\Gl_2,\Gl_3)$ involving matrix inverses. The
results are summarized in Theorem 1 at the end of the section.

We take as our Hilbert space $\CH$ those two-component periodic vector fields that are square integrable over the unit cell, with
the $L^2$-inner product of two fields, $\Ba(\Bx)$ and $\Bb(\Bx)$ being defined by 
\beq (\Ba,\Bb)=
( \bpm a_1\\ a_2 \epm, \bpm b_1 \\ b_2 \epm) =
\int_{\text{unit cell}} \overline{a_1({\Bx})}b_1({\Bx}) + \overline{a_2({\Bx})}b_2({\Bx})\,d{\Bx},
\eeq{34}
where the overline denotes complex conjugation, and we have adopted the (physicist's) convention of complex conjugating the fields
in the left of the inner product, rather than on the right.
Let 
\beqa
\CP_1 & = & \text{all vector fields in $\CH$ of the form }\bpm f_1(\Bx)\\ 0\epm,\nonum
\CP_2 & = & \text{all vector fields in $\CH$ of the form }\bpm 0\\ g_1(\Bx)\epm,\nonum
\CS & = & \text{all vector fields in $\CH$ of the form }\bpm f_2(\Bx)\\g_2(\Bx)\epm,
\eeqa{5}
with periodic functions $f_1(\Bx),g_1(\Bx),f_2(\Bx)$ and $g_2(\Bx)$ satisfying
$f_1(\Bx)\equiv g_1(\Bx)\equiv 0$ in phase $2$ and $f_2(\Bx)\equiv g_2(\Bx)\equiv 0$
in phase $1$.
Let 
\beqa
&&\BP_1\text{ denote the orthogonal projection onto $\CP_1$: $\BP_1 = \bpm 1 & 0 \\
0 & 0 \epm \Gc,$}\nonum
&&\BP_2\text{ denote the orthogonal projection onto $\CP_2$: $\BP_2 = \bpm 0 & 0 \\
0 & 1 \epm \Gc,$}\nonum
&&\BS\text{ denote the orthogonal projection onto $\CS$: $\BS = \bpm 1 & 0 \\
0 & 1 \epm (1-\Gc).$} \nonum
\eeqa{8}
Then we have
\beq \BP_1 + \BP_2 +\BS = \BI,\quad \BP_i^T = \BP_i,\quad \BS^T = \BS,\quad \BP_i\BP_j = \Gd_{ij}\BP_i,\quad \BP_i\BS=\BS\BP_i=0,
\eeq{12}
where $\BA^T$ denotes the adjoint operator of $\BA$ with respect to the inner product \eq{34},
and regarding the local conductivity $\BGs(\Bx)$ given by \eq{1} as an operator
acting on fields, we have 
\beq \BGs = \Gl_1\BP_1+\Gl_2\BP_2+\Gl_3\BS. \eeq{12a}
Let
\beqa
&&\CU_1=\text{the one-dimensional space of fields of the form $\bpm
e_1\\0\epm$},\nonum
&&\CU_2=\text{the one-dimensional space of fields of the form $\bpm
0\\e_2\epm$},
\eeqa{14}
where $e_1,e_2$ are constants independent of $\Bx$. Let
\beqa
\CE = \left\{\quad\begin{minipage}{0.8\textwidth}
curl-free fields which derive from periodic potentials,\\
i.e.\@ fields of the form $\bpm \frac{\partial \Gf}{\partial x_1}\\\frac{\partial \Gf}{\partial x_2}\epm$
for periodic $\Gf$,
\end{minipage}\right.\nonum
\CJ = \left\{\quad\begin{minipage}{0.8\textwidth}
divergence-free fields which derive from a periodic potentials,\\
i.e.\@ fields of the form $\bpm -\frac{\partial \Gy}{\partial x_2}\\ \frac{\partial \Gy}{\partial x_1}\epm$
for periodic $\Gy$.
\end{minipage}\right.
\eeqa{16}
Let
\beqa
&&\BGL_1\text{ denote the projection onto $\CU_1\oplus\CE$},\nonum
&&\BGL_2\text{ denote the projection onto $\CU_2\oplus\CJ$},
\eeqa{18}
which satisfy
\beq \BGL_1+\BGL_2=\BI,\quad \BGL_i^T = \BGL_i,\quad \BGL_i\BGL_j = \Gd_{ij}\BGL_i.
\eeq{21}
We then have the splitting of the space of square integrable vector
fields $\CH$ into orthogonal subspaces:
\beq
\CH = \CP_1\oplus\CP_2\oplus\CS = \CU_1\oplus\CU_2\oplus\CE\oplus\CJ,
\eeq{22}
where the spaces are orthogonal with respect to the usual $L^2$-inner
product \eq{34}.

Let $\BR_\perp$ denote the operator which locally rotates the fields by
$90^\circ$:
\beq
\BR_\perp \bpm h_1(\Bx)\\h_2(\Bx)\epm = \bpm 0 & -1\\ 1 & 0\epm \bpm
h_1(\Bx)\\h_2(\Bx)\epm = \bpm -h_2(\Bx)\\h_1(\Bx)\epm.
\eeq{23}
Of course we have $\BR_\perp^2 = -\BI$ and $\BR_\perp^T = -\BR_\perp$.
Note that 
\beqa
\BR_\perp\CU_1 = \CU_2, &\quad\BR_\perp \CE = \CJ,& \quad\BR_\perp\CJ = \CE,\\
\BR_\perp\CP_1 = \CP_2, &\quad\BR_\perp \CP_2 = \CP_1,&
\quad\BR_\perp\CS = \CS,
\eeqa{25}
or more specifically, the operators have the commutation properties
\beqa
\BR_\perp\BP_1 & = & \BP_2\BR_\perp, \quad\BR_\perp\BP_2 =
\BP_1\BR_\perp,\quad \BR_\perp\BS = \BS \BR_\perp,\nonum
\BR_\perp\BGG_0^{(1)} & = & \BGG_0^{(2)}\BR_\perp,\quad\BR_\perp\BGL_1=\BGL_2\BR_\perp,\quad\BR_\perp\BGL_2 = \BGL_1\BR_\perp,
\eeqa{27}
where $\BGG_0^{(i)}$ is the projection onto $\CU_i$ for $i=1,2$.

Let $\BGP$ be the operator which reflects a vector field about the $x_2$-axis.
Thus if $\Bg=\GP\Bh$ then the two components of $\Bg$ are related to the
two components of $\Bh$ via
\beq g_1(x_1,x_2)=h_1(-x_1,x_2), \quad  g_2(x_1,x_2)=-h_2(-x_1,x_2),
\eeq{27a}
This operator is self-adjoint, $\BGP^T=\BGP$, and
clearly commutes with $\BP_1$, $\BP_2$, $\BS$, $\BGL_1$, and $\BGL_2$:
\beq \BGP\BP_1=\BP_1\BGP,\quad \BGP\BP_2=\BP_2\BGP, \quad \BGP\BS=\BS\BGP,\quad \BGP\BGL_1=\BGL_1\BGP, \quad \BGP\BGL_2=\BGL_2\BGP,
\eeq{27b}  
and also anticommutes with $\BR_\perp$,
\beq \BGP\BR_\perp=-\BR_\perp\BGP, \eeq{27c}
since 
\beqa
\BGP\BR_\perp \bpm h_1(x_1,x_2)\\h_2(x_1,x_2)\epm & = & \BGP\bpm -h_2(x_1,x_2)\\h_1(x_1,x_2)\epm=
\bpm -h_2(-x_1,x_2)\\-h_1(x_1,x_2)\epm, \nonum
\BR_\perp \BGP\bpm h_1(x_1,x_2)\\h_2(x_1,x_2)\epm & = & \BR_\perp\bpm h_1(x_1,x_2)\\-h_2(x_1,x_2)\epm=
\bpm h_2(-x_1,x_2)\\h_1(x_1,x_2)\epm.
\eeqa{27d}
Note that $\BGP^2=\BI$ so the eigenvalues of $\BGP$ are either $+1$, corresponding to eigenfunctions
$\Bh^s(x_1,x_2)$ that are symmetric vector fields satisfying
\beq h_1^s(x_1,x_2)=h_1^s(-x_1,x_2), \quad  h_2^s(x_1,x_2)=-h_2^s(-x_1,x_2),
\eeq{27e}
or $-1$, corresponding to eigenfunctions
$\Bh^a(x_1,x_2)$ that are antisymmetric vector fields satisfying
\beq h_1^a(x_1,x_2)=-h_1^a(-x_1,x_2), \quad  h_2^a(x_1,x_2)=h_2^a(-x_1,x_2).
\eeq{27f}
Accordingly, we can define
\beqa
&&\CH^s=\text{all fields $\Bh^s\in\CH$ that satisfy \eq{27e}},\nonum
&&\CH^a=\text{all fields $\Bh^a\in\CH$ that satisfy \eq{27f}},
\eeqa{27fa}
and then $(\BI+\BGP)/2$ is the projection onto $\CH^s$, while $(\BI-\BGP)/2$ is the projection onto $\CH^a$.

Now the actual Hilbert space of interest, thus defined, is infinite-dimensional.  However, in the appendix
we show how this infinite-dimensional Hilbert space can be truncated to a finite-dimensional one with almost
no change to the effective conductivity function $\BGs^*(\BGs_1,\BGs_2)$, and with no change to the algebraic 
properties of the operators. So from now on we will assume the Hilbert space $\CH$ has finite dimension. 
Let us choose an orthonormal basis for $\CU_1\oplus\CE$:
\beq
\Bu_1,~\Bu_2,~\Bu_3,~\ldots,~\Bu_m.
\eeq{28}
We take then the fields
\beq
\Bv_1 = \BR_\perp\Bu_1,~\Bv_2 = \BR_\perp\Bu_2,~\ldots,~\Bv_m =
\BR_\perp\Bu_m,
\eeq{29}
as our basis for $\CU_2\oplus\CJ$. It follows that the $2m$ fields
$\Bu_1, \Bu_2, \ldots, \Bu_m, \Bv_1, \Bv_2, \ldots, \Bv_m$ form an
orthonormal basis for $\CH$. With respect to this basis we have
\beq
\BGL_1 = \bpm \BI & 0\\ 0 & 0\epm,\quad \BGL_2 = \bpm 0 & 0 \\ 0 & \BI\epm,
\quad\BR_\perp = \bpm 0 & -\BI\\ \BI & 0\epm,
\eeq{30}
where $\BI$ in each case is the $m\times m$ identity matrix.

Let us introduce the $m\times m$ matrices 
\beq V_{ij}^{(1)} = (\Bu_i,\BP_1\Bu_j),\quad V_{ij}^{(2)} = (\Bu_i,\BP_2\Bu_j),\quad W_{ij} = (\Bu_i,\BS\Bu_j),
\eeq{33}
for $i,j=1,2,\ldots,m$, and where the inner product of two fields is defined by \eq{34}.
These matrices then satisfy $\BV^{(1)}+\BV^{(2)}+\BW = \BI$. Using
the commutation properties of the operators we also have
\beqa
( \Bv_i,\BP_1\Bv_j )& = & ( \BR_\perp\Bu_i,\BP_1\BR_\perp\Bu_j) = (\Bu_i,\overbrace{\BR_\perp^T\BR_\perp}^{\BI}\BP_2\Bu_j) = V_{ij}^{(2)},\nonum
( \Bv_i,\BP_2\Bv_j ) & = & ( \BR_\perp\Bu_i,\BP_2\BR_\perp\Bu_j) = (\Bu_i,\BR_\perp^T\BR_\perp\BP_1\Bu_j) = V_{ij}^{(1)},\nonum
( \Bv_i,\BS\Bv_j ) & = &  ( \BR_\perp\Bu_i,\BS\BR_\perp\Bu_j) = (\Bu_i,\BR_\perp^T\BR_\perp\BS\Bu_j) = W_{ij}.
\eeqa{37}

Let us assume the basis $\Bu_1,\Bu_2,\ldots,\Bu_m$ has been chosen so
that $\BW$ is diagonal. Then we have
\beq
W_{ij} = \Gr_i\Gd_{ij}\implies \BW = \bpm\Gr_1 & 0 & \cdots
& 0\\
0 & \Gr_2 & \ddots & \vdots\\
\vdots & \ddots & \ddots & 0\\
0 & \cdots & 0 & \Gr_m\epm,
\eeq{38}
where (a priory) $0\leq \Gr_i \leq 1$, for $i=1,2\ldots,m$.

Equivalently, we choose each basis field $\Bu_i$ to be an eigenfield 
of the operator $\BGL_1\BS\BGL_1$, and then $\Gr_i$ is the corresponding eigenvalue.

Let $\rho$ denote an eigenvalue of the operator $\BGL_1\BS\BGL_1$, with $\rho\ne 0$ or $1$.
Let $\Be\ne 0$ denote a corresponding eigenfield, $\Be\in\CU_1\oplus\CE$. Then the complex
conjugate field $\overline\Be$ is also an eigenfield of $\BGL_1\BS\BGL_1$ with eigenvalue $\rho$.
By adding these two eigenfields, or subtracting $i$ times them, and relabeling the fields, we see that $\Be(\Bx)$ can be 
assumed to be real, without loss of generality. We have
\beq
\BGL_1\BS\Be = \Gr\Be.
\eeq{131}
Now consider the associated real-valued field 
\beq
\Be' = \BGL_1\BR_\perp\BS\Be\in\CU_1\oplus\CE.
\eeq{132}
This satisfies
\beqa
\BGL_1\BS\Be' &=& \BGL_1\BS\BGL_1\BR_\perp\BS\Be \nonum
&=&\BGL_1\BS\BR_\perp(\BI-\BGL_1)\BS\Be\quad\text{(since
$\BGL_1\BR_\perp=\BR_\perp\BGL_2$)} \nonum
&=&\BGL_1\BS\BR_\perp\BS\Be-\BGL_1\BS\BR_\perp\BGL_1\BS\Be \nonum
&=&\BGL_1\BR_\perp\BS\Be-\Gr\BGL_1\BR_\perp\BS\Be\quad\text{(since
$\BS\BR_\perp=\BR_\perp\BS$}\nonum
&\phantom{=}&\phantom{\BGL_1\BR_\perp\BS\Be-\Gr\BGL_1\BR_\perp\BS\Be\quad\text{(}}\text{and
$\BGL_1\BS\Be=\Gr\Be$) }\nonum
&=&(1-\Gr)\BGL_1\BR_\perp\BS\Be\nonum
&=&(1-\Gr)\Be'.
\eeqa{138}
Thus $\Be'$ is an eigenfield of $\BGL_1\BS\BGL_1$ with eigenvalue
$1-\Gr$. Note also that
\beqa
\BGL_1\BR_\perp\BS\Be'&=&\BGL_1\BR_\perp\BS\BGL_1\BR_\perp\BS\Be\nonum
&=&-\BGL_1\BS\BGL_2\BS\Be\nonum
&=&-\BGL_1\BS\Be+\BGL_1\BS\BGL_1\BS\Be\nonum
&=&-\Gr\Be+\Gr^2\Be = -\Gr(1-\Gr)\Be.
\eeqa{142}
Thus applying the operator $\BGL_1\BR_\perp\BS$ twice to $\Be$ brings one back 
to a multiple of $\Be$.
Note 
\beq
(\Be,\Be') = (\Be,\BR_\perp\BS\Be) = (\Be,\BS\BR_\perp\Be) =
-(\BR_\perp\BS\Be,\Be) = -(\Be',\Be).
\eeq{143}
So, as they are both real, $\Be$ and $\Be'$ must be orthogonal which implies $\Be'$ cannot be a
multiple of $\Be$. Then, using \eq{142}, we have
\beq
(\Be',\Be') = (\BGL_1\BR_\perp\BS\BGL_1\Be, \Be') =-(\Be,\BGL_1\BS\BR_\perp\Be')
=\Gr(1-\Gr)(\Be,\Be).
\eeq{146}
Since $\BW$ is the matrix representing the operator
$\BGL_1\BS\BGL_1$, it follows that the eigenvalues of $\BW$
occur in pairs $\Gr$ and $1-\Gr$.

Now since the self-adjoint operators $\BGP$ and $\BGL_1\BS\BGL_1$ commute, the basis 
fields $\Bu_1,\Bu_2,\ldots,\Bu_m$ can also be chosen to be eigenfunctions
of $\BGP$, i.e., either symmetric or antisymmetric fields. Let us suppose $\Be$ is 
a symmetric field so that $\BGP\Be=\Be$. Then using the commutation relations 
satisfied by $\BGP$, we see that $\Be'$ is an antisymmetric field:
\beq \BGP\Be'=\BGP\BGL_1\BR_\perp\BS\Be=-\BGL_1\BR_\perp\BS\BGP\Be=-\BGL_1\BR_\perp\BS\Be=-\Be'.
\eeq{38a}

Let us further assume that the basis fields are
ordered so the symmetric fields
come first. To simplify the analysis it is convenient to make the following assumption:
\vskip3mm
\noindent{\bf Assumption 1}
\vskip2mm
\noindent {\it Assume that none of the eigenvalues $\Gr_i$ defined by \eq{131} with $\Be\in\CU_1\oplus\CE$ and $\Be\ne 0$ are zero or one.} 
\vskip2mm

Thus we assume that a Hilbert 
space in which some of the eigenvalues $\Gr_i$ are zero or one can be 
approximated by one in which the eigenvalues $\Gr_i$ are not zero or one,
but rather close to zero, or close to one. It then follows that 
$m$ must be even since to every symmetric field $\Bu_i$ that is a
normalized eigenfunction of $\BGL_1\BS\BGL_1$ with eigenvalue $\Gr_i$ there is an associated 
antisymmetric field $\Bu'_i=\BGL_1\BR_\perp\BS\Bu_i/\sqrt{\Gr_i(1-\Gr_i)}$
that is a normalized eigenfunction of $\BGL_1\BS\BGL_1$ with eigenvalue $1-\Gr_i$,
and conversely
for every antisymmetric field $\Bu_i'$ that is a
normalized eigenfunction of $\BGL_1\BS\BGL_1$ with eigenvalue $1-\Gr_i$ there is an associated 
symmetric field $\Bu_i=-\BGL_1\BR_\perp\BS\Bu'_i/\sqrt{\Gr_i(1-\Gr_i)}$
that is a normalized eigenfunction of $\BGL_1\BS\BGL_1$ with eigenvalue $\Gr_i$. Thus
we can take as our basis the fields $\Bu_i$, which
are symmetric and orthonormal for $i\leq m/2$, and antisymmetric $i > m/2$,
with
\beq \Bu_{i+(m/2)}=\BGL_1\BR_\perp\BS\Bu_i/\sqrt{\Gr_i(1-\Gr_i)}, \eeq{38b}
for $i=1,2,\ldots, m/2$.

An argument similar to that in \eq{146} shows that for $i,j\leq m/2$
\beqa (\Bu_{j+(m/2)},\Bu_{i+(m/2)}) & = & (\BGL_1\BR_\perp\BS\BGL_1\Bu_j, \Bu_{i+(m/2)})/\sqrt{\Gr_j(1-\Gr_j)} \nonum
&= & -(\Bu_j,\BGL_1\BS\BR_\perp\Bu_{i+(m/2)})/\sqrt{\Gr_j(1-\Gr_j)} \nonum
& = &(\Bu_j,\Bu_i)\sqrt{\Gr_i(1-\Gr_i)}/\sqrt{\Gr_j(1-\Gr_j)}=\Gd_{ij},
\eeqa{38ba}
and consequently we see that the basis fields $\Bu_i$, $i=1,2,\ldots,m$ are all orthonormal (note that a symmetric field is automatically orthogonal to
an antisymmetric one). 

Now the fields $\Bv_i$ defined by \eq{29} have the property
that for $i\leq m/2$ they are antisymmetric fields:
\beq \BGP\Bv_i=\BGP\BR_\perp\Bu_i=-\BR_\perp\BGP\Bu_i=-\BR_\perp\Bu_i=-\Bv_i,
\eeq{38c}
while for $i> m/2$ they are symmetric fields:
\beq \BGP\Bv_i=\BGP\BR_\perp\Bu_i=-\BR_\perp\BGP\Bu_i=\BR_\perp\Bu_i=\Bv_i.
\eeq{38ca}
Consequently, in this basis the operator $\BGP$ has the block form:
\beq
\BGP = \bpm
\BI & 0 & 0 & 0\\
0 & -\BI & 0 & 0\\
0 & 0 & -\BI & 0\\
0 & 0 & 0 & \BI
\epm,
\eeq{38d}
while $\BGL_1$, $\BGL_2$, and $\BR_\perp$ clearly have the block form
\beqa
\BGL_1 &=& \bpm
\BI & 0 & 0 & 0\\
0 & \BI & 0 & 0\\
0 & 0 & 0 & 0\\
0 & 0 & 0 & 0
\epm,\nonum
\BGL_2 &=& \bpm
0 & 0 & 0 & 0\\
0 & 0 & 0 & 0\\
0 & 0 & \BI & 0\\
0 & 0 & 0 & \BI 
\epm,\nonum
\BR_\perp &=&\bpm
0 & 0 & -\BI & 0\\
0 & 0 & 0 & -\BI\\
\BI & 0 & 0 & 0 \\
0 & \BI & 0 & 0
\epm.
\eeqa{165}
If $\BU_1$ denotes a normalized field in $\CU_1$ (with $(\BU_1,\BU_1)=1$)
then since $\BU_1$ is a symmetric field satisfying $\BGL_1\BU_1=\BU_1$ it has 
the representation
\beq
\BU_1 = \bpm \BGb\\0 \\ 0 \\ 0\epm,\quad \BGb = \bpm
\Gb_1\\\Gb_2\\\vdots\\\Gb_{m/2}
\epm,
\eeq{165aa}
where, without loss of generality, we can assume that $\Gb_i\geq 0$ for all $i=1,2,\ldots,m/2$ (recall that we
can change the sign of $\Gb_i$ by changing the sign of the basis vector $\Bu_i$).

To obtain the matrix representing $\BS$ with respect to the basis
$\Bu_1,\Bu_2,\ldots,\Bu_m,\Bv_1,\Bv_2,\ldots,\Bv_m$ take $i,j\leq m/2$ and evaluate
\beqa (\Bv_{j+(m/2)},\BS\Bu_i)& = & -(\Bu_{j+(m/2)},\BGL_1\BR_\perp\BS\Bu_i) \nonum
                            & = & -(\Bu_{j+(m/2)},\Bu_{i+(m/2)})\sqrt{\Gr_i(1-\Gr_i)}=
-\Gd_{ij}\sqrt{\Gr_i(1-\Gr_i)}, \nonum
(\Bv_{j},\BS\Bu_i)& = &  -(\Bu_j,\Bu_{i+(m/2)})\sqrt{\Gr_i(1-\Gr_i)}=0,
\eeqa{165a}
and
\beqa (\Bv_j,\BS\Bu_{i+(m/2)}) & = & (\BGL_1\BS\BR_\perp\Bu_j,\Bu_{i+(m/2)}) \nonum
                             & = & (\Bu_{j+(m/2)},\Bu_{i+(m/2)})\sqrt{\Gr_j(1-\Gr_j)}
=\Gd_{ij}\sqrt{\Gr_i(1-\Gr_i)}, \nonum
(\Bv_{j+(m/2)},\BS\Bu_{i+(m/2)}) & = & (\BGL_1\BS\BR_\perp\Bu_{j+(m/2)},\Bu_{i+(m/2)}) \nonum
& = & -(\Bu_j,\Bu_{i+(m/2)})\sqrt{\Gr_j(1-\Gr_j)}=0.
\eeqa{165b}
Since the basis fields are real and $\BS$ is self-adjoint, it follows that
\beqa (\Bu_i,\BS\Bv_{j+(m/2)}) & = &(\Bv_{j+(m/2)},\BS\Bu_i)=-\Gd_{ij}\sqrt{\Gr_i(1-\Gr_i)},\nonum
(\Bu_i,\BS\Bv_j) & = & (\Bv_{j},\BS\Bu_i)=0, \nonum
(\Bu_{i+(m/2)},\BS\Bv_j) & = & (\Bv_j,\BS\Bu_{i+(m/2)})=\Gd_{ij}\sqrt{\Gr_i(1-\Gr_i)}, \nonum
(\Bu_{i+(m/2)},\BS\Bv_{j+(m/2)}) & = & (\Bv_{j+(m/2)},\BS\Bu_{i+(m/2)})=0.
\eeqa{165c}
From these formulae and from the last equations in \eq{33} and \eq{37} it follows that the matrix representing $\BS$ takes the form
\beq
\BS = \bpm 
\BZ_1 & 0 & 0 & -(\BZ_1\BZ_2)^{1/2}\\
0 & \BZ_2 & (\BZ_1\BZ_2)^{1/2} & 0\\
0 & (\BZ_1\BZ_2)^{1/2} & \BZ_1 & 0\\
-(\BZ_1\BZ_2)^{1/2} & 0 & 0 & \BZ_2
\epm,
\eeq{45}
in which $\BZ_2=\BI-\BZ_1$ and $\BZ_1$ is the diagonal matrix
\beq \BZ_1= \bpm\Gr_1 & 0 & \cdots
& 0\\
0 & \Gr_2 & \ddots & \vdots\\
\vdots & \ddots & \ddots & 0\\
0 & \cdots & 0 & \Gr_{m/2}\epm.
\eeq{45a}

Note that any field in $\CS$ is represented by a vector of the form
\beq
\bpm
\sqrt{\BZ_1}\Ba\\
\sqrt{\BZ_2}\Bb\\
\sqrt{\BZ_1}\Bb\\
-\sqrt{\BZ_2}\Ba
\epm,
\eeq{58}
where $\Ba$ and $\Bb$ are arbitrary $m/2$-component vectors. This
follows since such vectors are eigenvectors of $\BS$ with eigenvalue
$1$. Now what about the matrix representing $\BP_1$? The fact that
$\BP_1$ commutes with $\BGP$ implies that $\BP_1$ must have a similar
block structure (with eight zero block matrices positioned in the same places) as $\BS$. 
Also the fact that any vector of the form \eq{58}
must be a nullvector of $\BP_1$ for all $\Ba$ and $\Bb$ implies that $\BP_1$,
being a symmetric matrix, takes the form
\beq
\BP_1 = 
\left(\begin{array}{cc:cc}
\BY_1 & 0 & 0 & \BY_1\BQ\\
0 & \BY_2 & -\BY_2\BQ^{-1} & 0\\\hdashline
0 & -\BQ^{-1}\BY_2 & \BQ^{-1}\BY_2\BQ^{-1} & 0\\
\BQ\BY_1 & 0 & 0 & \BQ\BY_1\BQ
\end{array}\right),
\eeq{60}
where $\BQ= \sqrt{\BZ_1\BZ_2^{-1}}$ and $\BY_1,\BY_2$ are some symmetric
$(m/2)\times(m/2)$ matrices. Also, since $\BP_2 = \BR_\perp^T\BP_1\BR_\perp$ we have
\beq
\BP_2 = \bpm
\BQ^{-1}\BY_2\BQ^{-1} & 0 & 0 & \BQ^{-1}\BY_2\\
0& \BQ\BY_1\BQ & -\BQ\BY_1 & 0\\
0 & -\BY_1\BQ & \BY_1 & 0\\
\BY_2\BQ^{-1} & 0 & 0 & \BY_2
\epm.
\eeq{62}
Since $\BP_1+\BP_2 = \BI-\BS$ we have
\beqa
&&\BY_1 + \BQ^{-1}\BY_2\BQ^{-1} = \BI-\BZ_1 = \BZ_2,\nonum
&&\BY_2 + \BQ\BY_1\BQ = \BI - \BZ_2 = \BZ_1,\nonum
&&\BY_1\BQ+\BQ^{-1}\BY_2 = \sqrt{\BZ_1\BZ_2},\nonum
&&\BQ\BY_1+\BY_2\BQ^{-1}= \sqrt{\BZ_1\BZ_2}.
\eeqa{66}
These four equations are all equivalent and imply
\beq
\BY_2 = \BZ_1 - \BQ\BY_1\BQ.
\eeq{67}
We also have the equation $\BP_1^2 = \BP_1$. Note that
\beq
\BP_1^2 = 
\left(
\renewcommand\arraystretch{1.5}
\begin{array}{c:c}
\wt{\BP}_{11} & \wt{\BP}_{12}\\\hdashline
\wt{\BP}_{12}^T & \wt{\BP}_{22}
\end{array}\right),
\eeq{68}
where
\beqa
\wt{\BP}_{11}&=&\bpm
\BY_1(\BI+\BQ^2)\BY_1 & 0  \\
0 & \BY_2(\BI+\BQ^{-2})\BY_2 
\epm,\nonum
\wt{\BP}_{12}&=&\bpm
0 & \BY_1(\BI+\BQ^2)\BY_1\BQ\\
-\BY_2(\BI+\BQ^{-2})\BY_2\BQ^{-1} & 0
\epm,\nonum
\wt{\BP}_{22}&=&\bpm
\BQ^{-1}\BY_2(\BQ^{-2}+\BI)\BY_2\BQ^{-1} & 0\\
 0 & \BQ\BY_1(\BI+\BQ^2)\BY_1\BQ
\epm.
\eeqa{68a}
Equating this with $\BP_1$ gives the equations
\beq
\BY_1 = \BY_1(\BI+\BQ^2)\BY_1,\quad \BY_2 = \BY_2(\BI+\BQ^{-2})\BY_2.
\eeq{69}
Let
\beqa
\BGU_1\text{ denote the projection onto the range $\CR_1$ of $\BY_1$,}\nonum
\BGU_2\text{ denote the projection onto the range $\CR_2$ of $\BY_2$}.
\eeqa{71}
Since $\BY_1$ and $\BY_2$ are symmetric they commute with $\BGU_1$ and $\BGU_2$ respectively
Then we have
\beq
\BY_1 = \BY_1(\BGU_1(\BI+\BQ^2)\BGU_1)\BY_1,\quad
\text{i.e.\@ } \BY_1 = \BGU_1(\BGU_1(\BI+\BQ^2)\BGU_1)^{-1}\BGU_1,
\eeq{73}
where the inverse is to be taken on the space $\CR_1$. Similarly, we
find
\beq
\BY_2 =\BGU_2(\BGU_2(\BI+\BQ^{-2})\BGU_2)^{-1}\BGU_2,
\eeq{74}
where the inverse is to be taken on the space $\CR_2$. In other words,
if we know $\BGU_1$ and $\BGU_2$, we can determine both $\BY_1$ and
$\BY_2$. We also have $\BP_1\BP_2=0$, i.e. 
\beq
\BP_1\BP_2 = 
\renewcommand\arraystretch{1.5}
\left(\begin{array}{c:c}
\BP_{11}' & \BP_{12}'\\\hdashline
\BP_{21}' & \BP_{22}'
\end{array}\right)= 0,
\eeq{75}
where
\beqa
\BP_{11}' &=&
\bpm
\BY_1(\BQ^{-1}+\BQ)\BY_2\BQ^{-1} & 0\\
0      & \BY_2(\BQ^{-1}+\BQ)\BY_1\BQ
\epm,\nonum
\BP_{12}' &=&
\bpm
0 & \BY_1(\BQ^{-1}+\BQ)\BY_2\\
-\BY_2(\BQ^{-1}+\BQ)\BY_1 & 0
\epm,\nonum
\BP_{21}' &=&
\bpm
0 & -\BQ^{-1}\BY_2(\BQ+\BQ^{-1})\BY_1\BQ\\
\BQ\BY_1(\BQ+\BQ^{-1})\BY_2\BQ^{-1}& 0
\epm,\nonum
\BP_{22}' &=& 
\bpm
\BQ^{-1}\BY_2(\BQ+\BQ^{-1})\BY_1 & 0\\
 0 & \BQ\BY_1(\BQ+\BQ^{-1})\BY_2
\epm.
\eeqa{75a}
Equality with $0$ will be satisfied provided 
\beq
\BY_1(\BQ+\BQ^{-1})\BY_2 =
0.
\eeq{76}
This relation is in fact implied by the relations
\beq
\BY_1 = \BY_1(\BI+\BQ^2)\BY_1,\quad \BZ_1+\BZ_2-\BI = 0,\quad \BY_2 =
\BZ_1-\BQ\BY_1\BQ.
\eeq{77}
To see this note that
\beqa
\BY_1(\BQ+\BQ^{-1})\BY_2 &=& \BY_1(\BQ+\BQ^{-1})(\BZ_1-\BQ\BY_1\BQ)\nonum
&=&\BY_1\BQ\BZ_1+\BY_1\BQ^{-1}\BZ_1-\BY_1(\BI+\BQ^2)\BY_1\BQ\nonum
&=&\BY_1\BQ\BZ_1+\BY_1\BQ\BZ_2-\BY_1\BQ\nonum
&=&\BY_1\BQ(\BZ_1+\BZ_2-\BI)=0.
\eeqa{81}
It remains to show that the relation
\beq
\BY_2 = \BY_2(\BI+\BQ^{-2})\BY_2
\eeq{82}
is implied by the identities
\beq
\BY_2 =\BZ_1 - \BQ\BY_1\BQ, \quad\BY_1 = \BY_1(\BI+\BQ^2)\BY_1,\quad \BZ_1+\BZ_2 =
\BI.
\eeq{83}
We have
\beqa
\BY_2 - \BY_2(\BI+\BQ^{-2})\BY_2 &=& \BY_2 -
\BY_2(\BI+\BQ^{-2})(\BZ_1-\BQ\BY_1\BQ)\nonum
&=& \BY_2 - \BY_2(\BI +\BQ^{-2})\BZ_1+\BY_2(\BQ+\BQ^{-1})\BY_1\BQ\nonum
&=& \BY_2(\BI-\BZ_1-\BZ_2)\nonum
&=&0.
\eeqa{87}
In summary, if we are given $\BGU_1$ then we take
\beq
\BY_1 = (\BGU_1(\BI+\BQ^2)\BGU_1)^{-1},\quad
\BY_2 = \BZ_1 - \BQ\BY_1\BQ.
\eeq{89}
Also note that
\beq
\BI+\BQ^2 = \BI + \BZ_1\BZ_2^{-1} = (\BZ_2+\BZ_1)\BZ_2^{-1} =
\BZ_2^{-1}.
\eeq{90}

Let $n=\rank(\BY_1)$. If we assume $\BZ_1$ is strictly positive
definite, then
\beq
\BY_2 + \BQ\BY_1\BQ = \BZ_1 = \text{a positive definite
$\frac{m}{2}\times\frac{m}{2}$ matrix}
\eeq{91}
is non-singular. Hence $\BY_2$ must have rank at least $m/2 - n$. Also
since
\beq
\BY_2(\BQ+\BQ^{-1})\BY_1 = 0,
\eeq{92}
the nullspace of $\BY_2$ must be at least of dimension $n$. Putting
these facts together we see that
\beq
\rank(\BY_2) = \frac{m}{2}-n.
\eeq{93}

As we are trying to treat the generic case, let us make for simplicity
the further assumption:
\newpage
\vskip3mm
\noindent{\bf Assumption 2}
\vskip2mm
\noindent{\it We assume the fields}
\beq
\Bw_1=\BGU_1\Bu_1, \quad\Bw_2=\BGU_1\Bu_2, \quad\ldots,\quad \Bw_n = \BGU_1\Bu_n,
\eeq{94}
{\it are non-zero and independent, where $\BGU_1$ is the projection onto the range of $\BY_1$ which is defined as that operator $(\BI+\BGP)\BGL_1\BP_1\BGL_1(\BI+\BGP)/4$
appearing in the representation \eq{60} of $\BP_1$,
and the $\Bu_i\in\CH^s\cap(\CU_1\oplus\CE)$, specified following Assumption 1, are orthonormal eigenfields of $\BGL_1\BS\BGL_1$. } 
\vskip2mm

We can then take these as a basis for
$\CR_1$. With respect to this basis, the action of $\BGU_1$ on a field
$\Bu_i$ for $i=1,2,\ldots,m/2$, is represented by a matrix $\BK$:
\beq
\BGU_1\Bu_i = \sum_{a=1}^n \Bw_aK_{ai},
\eeq{95}
where $\BK$ is an $n\times(m/2)$ matrix. The definition of the
basis implies $K_{ai} = \Gd_{ai}$ for all $a=1,\ldots,n,$ when
$i=1,\ldots,n,$ but not when $i=n+1,n+2,\ldots, m/2$. In other words,
the $n\times\frac{m}{2}$ matrix $\BK$ has the special form
\beq
\BK = \bpm \BI & \BH\epm,
\eeq{119}
in which $\BI$ is the $n\times n$ identity matrix and $\BH$ is an
$n\times(\frac{m}{2}-n)$ matrix. By taking the
inner product with respect to $\Bu_b$ for $b=1,2,\ldots,n$, we deduce
that
\beq
( \Bu_b,\BGU_1\Bu_i) =
\sum_{a=1}^n(\Bu_b,\BGU_1\Bu_a) K_{ai}= \sum_{a=1}^n M_{ba}K_{ai},
\eeq{97}
where $\BM$ is the $n\times n$ symmetric positive definite matrix with
elements $M_{ba}=( \Bu_b,\BGU_1\Bu_a)$. Since the $\Bu_i$'s
form an orthonormal basis it follows that
\beq
\Bw_a = \BGU_1\Bu_a = \sum_{j=1}^{m/2}( \Bu_a,\BGU_1\Bu_j)\Bu_j=\sum_{j=1}^{m/2}\sum_{c=1}^nM_{ac}K_{cj}\Bu_j,
\eeq{99}
for $a=1,2,\ldots,n$. Hence we obtain
\beq
\BGU_1\Bu_i = \sum_{a=1}^n\Bw_aK_{ai} =
\sum_{a=1}^n\sum_{j=1}^{m/2}\sum_{c=1}^nK_{ai}M_{ac}K_{cj}\Bu_j,
\eeq{100}
i.e. $\BGU_1$ is represented by the matrix
\beq
\BGU_1 = \BK^T\BM\BK,
\eeq{101}
where $\BM$ is symmetric and $\BK$ is of the form
\beq
\BK = \bpm \BI & \BH\epm\quad\text{with } \BH\text{ an 
$n\times(m/2-n)$ matrix.}
\eeq{102}

Since $\BGU_1^2 = \BGU_1$, we have
\beq
\BK^T\BM\BK\BK^T\BM\BK = \BK^T\BM\BK,
\eeq{103}
i.e. $\BK\BK^T = \BM^{-1}$ or $\BM= (\BK\BK^T)^{-1}.$ Therefore we have
\beq
\BGU_1 = \BK^T\BM\BK\quad\text{where } \BM=(\BK\BK^T)^{-1}\text{ and } \BK =
\bpm \BI & \BH\epm.
\eeq{104}
Thus to determine $\BGU_1$ all we need to specify is the matrix $\BH$.
Note the number of parameters in $\BH$ is
\beq
k(n) = n(m/2-n) = \frac{mn}{2}-n^2.
\eeq{105}
In the worst case scenario, with $m$ fixed, we treat $n$ as a continuous variable
and find the value of $n$ which maximizes $k(n)$:
\beq
\frac{dk(n)}{dn} = \frac{m}{2}-2n = 0\quad\text{when }n=\frac{m}{4}. 
\eeq{106}
Thus the worst scenario is when $\BH$ is nearly square, in which case the number
of parameters is
\beq
k(n) \approx \left(\frac{m}{4}\right)\left(\frac{m}{4}\right) = \frac{m^2}{16}.
\eeq{107}
The fields $\Bw_i$ are not orthonormal. From \eq{99} we have
\beqa
(\Bw_a,\Bw_b) & = & \sum_{j=1}^{m/2}\sum_{c=1}^n\sum_{i=1}^{m/2}\sum_{d=1}^nM_{ac}K_{cj}M_{bd}K_{di}(\Bu_j,\Bu_i) \nonum
& = & \sum_{j=1}^{m/2}\sum_{c=1}^n\sum_{d=1}^n \{\BM\BK\BK^T\BM\}_{ab}=M_{ab}.
\eeqa{107a}
So we may take the fields
\beq \Bw_a'=\sum_b\{\BM^{-1/2}\}_{ab}\Bw_b, \eeq{107b}
as our orthonormal basis. This and \eq{99} imply
\beq
\Bw_a' =\sum_{j=1}^{m/2}\sum_{c=1}^n\{\BM^{1/2}\BK\}_{aj}\Bu_j.
\eeq{107c}
With respect to this basis the operator $\BGU_1(\BI+\BQ^2)\BGU_1=\BGU_1\BZ_2^{-1}\BGU_1$
is represented by the matrix
\beqa \{\BGU_1\BZ_2^{-1}\BGU_1\}_{ab} & = &(\Bw_a',\BZ_2^{-1}\Bw_b')=\sum_{j=1}^{m/2}\sum_{i=1}^{m/2}\{\BM^{1/2}\BK\}_{aj}\{\BM^{1/2}\BK\}_{bi}(\Bu_j,\BZ_2^{-1}\Bu_i)\nonum
& = & \{\BM^{1/2}\BK\BZ_2^{-1}\BK^T\BM^{1/2}\}_{ab}.
\eeqa{107d}
Given $\BH$ (and hence $\BK$ and $\BM$) the formula \eq{89} for $\BY_1$ (in the basis $\Bu_i$) is therefore
\beqa
\BY_1 &=&
\BK^T\BM^{1/2}(\BM^{1/2}\BK\BZ_2^{-1}\BK^T\BM^{1/2})^{-1}\BM^{1/2}\BK\nonum
&=&\BK^T(\BK\BZ_2^{-1}\BK^T)^{-1}\BK,
\eeqa{110}
and
\beq
\BY_2 = \BZ_1-\BQ\BY_1\BQ,
\eeq{111}
where $\BQ=(\BZ_1\BZ_2^{-1})^{1/2}$ and $\BZ_2 = \BI-\BZ_1$.
%
Then we have
\beq
\BV^{(1)} = \bpm \BY_1 & 0 \\ 0 & \BY_2\epm,\quad \BV^{(2)} =
\bpm(\BQ^{-1}\BY_2\BQ^{-1} & 0 \\ 0 & \BQ\BY_1\BQ\epm,\quad \BW =
\bpm\BZ_1 & 0 \\ 0 & \BZ_2\epm.
\eeq{112}
To obtain a formula for the diagonal element $\Gs^*_{11}$ of the 
effective conductivity tensor $\BGs^*$, we start from the observation
that 
\beq \lang j_1\rang=\BU_1\cdot\Bj=\BU_1\cdot(\BGL_1\Bj)=\BU_1\cdot(\BGL_1\BGs\Be)=\BU_1\cdot(\BGL_1\BGs\BGL_1)\Be,
\eeq{112a}
where $\BU_1$, with $(\BU_1\cdot\BU_1)=1$, is in $\CU_1$ and can be represented by a column vector of the form $\bpm \Bu_0\\ 0\epm$, thus defining $\Bu_0$.
This implies that 
$\Be=(\BGL_1\BGs\BGL_1)^{-1}\BU_1\lang j_1\rang$, where the operator inverse is to be taken on the space $\CU_1\oplus\CE$. Hence
\beq \lang e_1\rang=\BU_1\cdot\Be=\BU_1\cdot(\BGL_1\BGs\BGL_1)^{-1}\BU_1\lang j_1\rang,
\eeq{112b}
and this gives the formula
\beq (\Gs^*_{11})^{-1}=\BU_1\cdot(\BGL_1\BGs\BGL_1)^{-1}\BU_1. \eeq{112c}

Since $\BV^{(1)},\BV^{(2)}$ and $\BW$ are all block diagonal, we have that 
the diagonal element $\Gs^*_{11}(\Gl_1,\Gl_2,\Gl_3)$ of the effective conductivity tensor,
that determines the tensor function $\BGs^*(\Gl_1,\Gl_2,\Gl_3)$  through \eq{2.a} and \eq{2.b}
is going to be given by the formula
\beqa
[\Gs^*_{11}(\Gl_1,\Gl_2,\Gl_3)]^{-1} &=&
\Bu_0\cdot(\BV^{(1)}\Gl_1+\BV^{(2)}\Gl_2+\BW\Gl_3)^{-1}\Bu_0\nonum
&=&
\BGb\cdot(\BY_1\Gl_1+\underbrace{\BQ^{-1}\BY_2\BQ^{-1}\Gl_2}_{\BZ_2-\BY_1}+\BZ_1\Gl_3)^{-1}\cdot\BGb\\
&=&\BGb\cdot(\BZ_2\Gl_2+\BZ_1\Gl_3+\BY_1(\Gl_1-\Gl_2))^{-1}\cdot\BGb.
\eeqa{115}
where $\Bu_0$, defined just after \eq{112a}, has the representation $\bpm \BGb\\ 0\epm$ which defines $\BGb$: see \eq{165aa}. This formula
is exact on the truncated Hilbert space, but only approximate on the original physical Hilbert space.

In summary, we have the following Theorem:
\vskip3mm
\noindent{\bf Theorem 1}
\vskip2mm
{\it Suppose the conductivity tensor has the form \eq{1}, and consider the Domain $\CD(c_1,c_2)$ introduced in the appendix of pairs $(\BGs_1,\BGs_2)$ 
such that the corresponding triplet $(\Gl_1,\Gl_2,\Gl_3)$ satisfies
\beq c_1\leq \Real(\Gl_i),\quad |\Gl_i|\leq c_2,\quad i=1,2,3,
\eeq{115aa}
where $c_1$, $c_2$ are fixed real constants with $c_2>c_1>0$. 
Subject to Assumptions 1 and 2, the diagonal element $\Gs^*_{11}(\Gl_1,\Gl_2,\Gl_3)$ of the effective conductivity tensor $\BGs_*$
can be approximated arbitrarily closely for $(\BGs_1,\BGs_2)\in\CD(c_1,c_2)$ by
\beq
[\Gs^*_{11}(\Gl_1,\Gl_2,\Gl_3)]^{-1} \approx
\BGb\cdot(\BZ_2\Gl_2+\BZ_1\Gl_3+\BY_1(\Gl_1-\Gl_2))^{-1}\BGb,
\eeq{116}
where $\BZ_1,\BZ_2=\BI-\BZ_1$ are diagonal positive definite $\frac{m}{2}\times\frac{m}{2}$
matrices, $\BGb$ is an $m/2$-component vector with non-negative
entries, and the $\frac{m}{2}\times\frac{m}{2}$ matrix $\BY_1$ takes the form
\beq
\BY_1 = \BK^T(\BK\BZ_2^{-1}\BK^T)^{-1}\BK,
\eeq{118}
where the $n\times\frac{m}{2}$ matrix $\BK$ has the special form
\beq
\BK = \bpm \BI & \BH\epm,
\eeq{119a}
in which $\BI$ is the $n\times n$ identity matrix and $\BH$ is an
$n\times(\frac{m}{2}-n)$ matrix.}
\vskip2mm

When $\Gl_1=\Gl_2=1$ and $\Gl_3=\Gl$ we have 
\beqa
[\Gs^*_{11}(1,1,\Gl)]^{-1} &=& \Bu_0\cdot\bpm \Gl\Gr_1+(1-\Gr_1) & 0 &
\cdots & 0\\
0 & \Gl\Gr_2+(1-\Gr_2) & \ddots & \vdots\\
\vdots & \ddots & \ddots & \vdots\\
0 & \cdots & 0 & \Gl\Gr_m+(1-\Gr_m)
\epm^{-1}\Bu_0\nonum
&=& \sum_{i=1}^{m/2}\frac{\Gb_i^2}{\Gl\Gr_i+(1-\Gr_i)},
\eeqa{48}
where $\Bu_0 = (\Gb_1,\Gb_2,\cdots,\Gb_{m/2},0,\ldots,0)^T$. Assuming 
none of the $\Gb_i$ are zero for $i\leq m/2$ we can determine from the 
poles of $[\Gs^*_{11}(1,1,\Gl)]^{-1}$ the parameters $\Gr_i$, and 
hence the matrices $\BZ_1$ and $\BZ_2$, and from the residues we can determine the
parameters $\Gb_i$.

An obvious but challenging question is whether, in the generic case, we
can recover the parameters of the matrix $\BH$, in addition to the 
parameters $\Gb_i$ and $\Gr_i$ from knowledge of the function 
$[\Gs^*_{11}(\Gl_1,\Gl_2,\Gl_3)]^{-1}$ given by \eq{116}.

\section{Representations for the operators when both phases are anisotropic}
\setcounter{equation}{0}
If both phases are anisotropic then we need to use the spaces
\beqa
\CP_1 & = & \text{all vector fields of the form }\bpm f_1(\Bx)\\ 0\epm,\nonum
\CP_2 & = & \text{all vector fields of the form }\bpm 0\\ g_1(\Bx)\epm,\nonum
\CP_3 & = & \text{all vector fields of the form }\bpm f_2(\Bx)\\ 0\epm,\nonum
\CP_4 & = & \text{all vector fields of the form }\bpm 0\\ g_2(\Bx)\epm,\nonum
\eeqa{149a}
with periodic functions $f_1(\Bx),g_1(\Bx),f_2(\Bx)$ and $g_2(\Bx)$ satisfying
$f_1(\Bx)\equiv g_1(\Bx)\equiv 0$ in phase $2$ and $f_2(\Bx)\equiv g_2(\Bx)\equiv 0$
in phase $1$.
Accordingly, we let
\beqa
&&\BP_1\text{ denote the orthogonal projection onto } \CP_1: \BP_1 = \bpm 1 & 0 \\
0 & 0 \epm \Gc,\nonum
&&\BP_2\text{ denote the orthogonal projection onto } \CP_2: \BP_2 = \bpm 0 & 0 \\
0 & 1 \epm \Gc,\nonum
&&\BP_3\text{ denote the orthogonal projection onto } \CP_3: \BP_3 = \bpm 1 & 0 \\
0 & 0 \epm(1-\Gc),\nonum
&&\BP_4\text{ denote the orthogonal projection onto } \CP_4: \BP_4 = \bpm 0 & 0 \\
0 & 1 \epm (1-\Gc). \nonum
\eeqa{150}
Note that Assumption 1 in the previous section, that the
operator $\BGL_1(\BP_3+\BP_4)\BGL_1$ on the space $\CU_1\oplus\CE$ does not have $0$ or $1$ as an eigenvalue,
implies that  
\beq  \BGL_1(\BP_1+\BP_2)\BGL_1=\BGL_1(\BI-\BP_3-\BP_4)\BGL_1=\BGL-\BGL_1(\BP_3+\BP_4)\BGL_1,
\eeq{150cc}
also does not have 0 or 1 as an eigenvalue. So by a similar analysis as in the previous section, and after 
an appropriate truncation of the Hilbert space, such as that described in the Appendix, 
these operators have the representations
\beqa
\BP_1 &=& \bpm
\BY_1 & 0 & 0 & \BY_1\BQ\\
0 & \BY_2 & -\BY_2\BQ^{-1} & 0\\
0 & -\BQ^{-1}\BY_2 & \BQ^{-1}\BY_2\BQ^{-1} & 0\\
\BQ\BY_1 & 0 & 0 & \BQ\BY_1\BQ\\
\epm,\nonum
\BP_2 &=& \bpm
\BQ^{-1}\BY_2\BQ^{-1} & 0 & 0 & \BQ^{-1}\BY_2\\
0 & \BQ\BY_2\BQ & -\BQ\BY_1 & 0\\
0 & -\BY_1\BQ & \BY_1 & 0\\
\BY_2\BQ^{-1} & 0 & 0 & \BY_2
\epm,\nonum
\BP_3 &=& \bpm
\BY_3 & 0 & 0 & \BY_3\BQ^{-1}\\
0 & \BY_4 & -\BY_4\BQ & 0\\
0 & -\BQ\BY_4 & \BQ\BY_4\BQ & 0\\
\BQ^{-1}\BY_3 & 0 & 0 & \BQ^{-1}\BY_3\BQ^{-1}
\epm,\nonum
\BP_4 &=& \bpm
\BQ\BY_4\BQ & 0 & 0 & \BQ\BY_4\\
0 & \BQ^{-1}\BY_3\BQ^{-1} & -\BQ^{-1}\BY_3 & 0\\
0 & -\BY_3\BQ^{-1} & \BY_3 & 0\\
\BY_4\BQ & 0 & 0 & \BY_4
\epm.
\eeqa{154}
Here, as before, we obtain
\beq \BQ = \BZ_1^{1/2}\BZ_2^{-1/2},\quad \BZ_2 = \BI-\BZ_1,\quad
\BZ_1 = \bpm
\Gr_1 & 0 & \cdots & 0\\
0 & \Gr_2 & \ddots & \vdots\\
\vdots & \ddots & \ddots & 0\\
0 & \cdots & 0 & \Gr_{m/2}
\epm.
\eeq{157}

Making Assumption 2, with $\BS$ replaced by $\BP_3+\BP_4$, and making
the analogous assumption:
\vskip3mm
\noindent{\bf Assumption 3}
\vskip2mm
\noindent{\it We assume the fields}
\beq
\Bw_1=\BGU_3\Bu_1, \quad\Bw_2=\BGU_3\Bu_2, \quad\ldots,\quad \Bw_n = \BGU_3\Bu_n,
\eeq{157aa}
{\it are non-zero and independent, where $\BGU_3$ is the projection onto the range of $\BY_3$ which is defined as the operator 
$(\BI+\BGP)\BGL_1\BP_3\BGL_1(\BI+\BGP)/4$ appearing in the representation \eq{157} of $\BP_3$,
 and the $\Bu_i\in\CH^s\cap(\CU_1\oplus\CE)$, specified following Assumption 1, are orthonormal eigenfields of $\BGL_1(\BP_3+\BP_4)\BGL_1$, 
which are the same as the eigenfields of $\BGL_1(\BP_1+\BP_2)\BGL_1$.} 

\vskip2mm

\noindent we then have
\beqa
\BY_1 = \BK_1^T(\BK_1\BZ_2^{-1}\BK_1^T)^{-1}\BK_1,&&\BY_2 =
\BZ_1-\BQ\BY_1\BQ,\nonum
\BY_3 = \BK_2^T(\BK_2\BZ_1^{-1}\BK_2^T)^{-1}\BK_2 && \BY_4 = \BZ_2 -
\BQ^{-1}\BY_3\BQ^{-1},
\eeqa{159}
where
\beqa
&\BK_1 = (\BI \quad\BH_1): \text{$\BI$ is the $n_1\times n_1$ identity,
$\BH_1$ is $n_1\times(\frac{m}{2}-n_1)$},\nonum
&\BK_2 = (\BI \quad \BH_2): \text{$\BI$ is the $n_2\times n_2$ identity,
$\BH_2$ is $n_2\times(\frac{m}{2}-n_2)$.}
\eeqa{161}
The remaining operators $\BGP$, $\BGL_1$, $\BGL_2$ and $\BR_\perp$ are defined as before,
and have the same representations \eq{38d} and \eq{165} as before. Also $\BU_1$ still
has the representation \eq{165aa} in terms of the parameters $\Gb_i$.

Note that since $\BU_1$ lies in the space $\CP_1\oplus\CP_3$ we should have $(\BP_1+\BP_3)\BU_1=\BU_1$, i.e.,
\beq
(\BP_1+\BP_3)\BU_1 =\bpm(\BY_1+\BY_3)\BGb\\ 0 \\ 0
\\(\BQ\BY_1+\BQ^{-1}\BY_3)\BGb\epm = \bpm
\BGb\\0\\0\\0\epm.
\eeq{166}
Equivalently, $\BU_1$ should be a
nullvector of $\BP_2$ and $\BP_4$. That is
\beq
\BQ^{-1}\BY_2\BQ^{-1}\BGb = 0 \text{ and } \BQ\BY_4\BQ\BGb =
0.
\eeq{167}
Since $\BQ^{-1}\BY_2\BQ^{-1} = \BZ_2 - \BY_1$ and $\BQ\BY_4\BQ = \BZ_1 -\BY_3$, we deduce that
\beq
\BY_1\BGb = \BZ_2\BGb\text{ and } \BY_3\BGb =
\BZ_1\BGb.
\eeq{168}
An alternative way of expressing this is to say there exist vectors
$\Bv_1$ and $\Bv_2$ such that 
\beq
\BK_1^T\Bv_1 = \BZ_2\BGb\text{ and } \BK_2^T\Bv_2 =
\BZ_1\BGb.
\eeq{169}
To see the equivalence notice that if $\BK_1^T\Bv_1=\BZ_2\BGb$,
then
\beqa
\BY_1\BGb &=&\BK_1^T(\BK_1\BZ_2^{-1}\BK_1^T)^{-1}\BK_1\BGb\nonum
&=&\BK_1^T(\BK_1\BZ_2^{-1}\BK_1^T)^{-1}\BK_1\BZ_2^{-1}\BK_1^T\Bv_1\quad\text{(since
$\BGb=\BZ_2^{-1}\BK_1^T\Bv_1$)}\nonum
&=&\BK_1^T\Bv_1\nonum
&=&\BZ_2\BGb.
\eeqa{173}
Similarly, if $\BK_2^T\Bv_2=\BZ_1\BGb$ then $\BY_3\BGb =
\BZ_1\BGb$.

The equation $\BK_1^T\Bv_1=\BZ_2\BGb$ can be written as
\beq
\bpm\BI\\\BH_1^T\epm\Bv_1 = \BZ_2\BGb = \bpm
(1-\Gr_1)\Gb_1\\(1-\Gr_2)\Gb_2\\\vdots\\(1-\Gr_{m/2})\Gb_{m/2}\epm,
\eeq{174}
implying that
\beq
\Bv_1 = \bpm
(1-\Gr_1)\Gb_1\\(1-\Gr_2)\Gb_2\\\vdots\\(1-\Gr_{n_1})\Gb_{n_1}\epm,
\eeq{175}
and
\beq
\BH_1^T\bpm(1-\Gr_1)\Gb_1\\(1-\Gr_2)\Gb_2\\\vdots\\(1-\Gr_{n_1})\Gb_{n_1}\epm=\bpm(1-\Gr_{n_1+1})\Gb_{n_1+1}\\(1-\Gr_{n_1+2})\Gb_{n_1+2}\\\vdots\\(1-\Gr_{m/2})\Gb_{m/2}\epm.
\eeq{176}
This equation can be generally used (when $(1-\Gr_{m/2})\Gb_{m/2}$ is nonzero) to determine the last column of the matrix
$\BH_1^T$ given the first $n_1-1$ columns. In other words, it can be
used to determine the last row of the matrix $\BH_1$ given the first
$n_1-1$ rows. Similarly, we have
\beq
\Bv_2 = \bpm \Gr_1\Gb_1\\\Gr_2\Gb_2\\\vdots\\\Gr_{n_2}\Gb_{n_2}\epm,
\eeq{177}
and
\beq
\BH_2^T\bpm\Gr_1\Gb_1\\\Gr_2\Gb_2\\\vdots\\\Gr_{n_2}\Gb_{n_2}\epm = \bpm
\Gr_{n_2+1}\Gb_{n_2+1}\\\Gr_{n_2+2}\Gb_{n_2+2}\\\vdots\\\Gr_{m/2}\Gb_{m/2}\epm.
\eeq{178}
Again, this equation can be generally used (when $\Gr_{n_2}\Gb_{n_2}$ is nonzero) to determine the last column of the matrix
$\BH_2^T$ given the first $n_2-1$ columns. In other words, it can be
used to determine the last row of the matrix $\BH_2$ given the first
$n_2-1$ rows.

\section{An approximate formulae for the effective matrix-valued effective conductivity function with possibly nondiagonal, nonsymmetric, conductivity tensors of the phases.}
\setcounter{equation}{0}
Here we seek an approximate representation formula for the function $\BGs_*(\BGs_1,\BGs_2)$ allowing the matrices $\BGs_1$ and $\BGs_2$ to be nondiagonal and even not necessarily
symmetric (although the case where they are nonsymmetric can be reduced to that where they are symmetric \cite{Milton:1988:CHE}, see also chapter 4 in \cite{Milton:2002:TOC}). The results are summarized at the end of the section in Theorem 2.

Allowing for general anisotropies of the two conductivity tensors $\BGs_1$ and $\BGs_2$ of the two
phases, we can express the conductivity operator $\BGs(\Bx)$ in terms of the operators
$\BP_1$, $\BP_2$, $\BP_3$, $\BP_4$ and $\BR_\perp$ for which we have representation formulas:
explicitly, in direct analogy with equation (2.25) in \cite{Clark:1994:MEC} we have
\beqa
\BGs(\Bx) &=& \BGs_1\Gc+\BGs_2(1-\Gc)=
\bpm \Gs_{1,11} & \Gs_{1,12} \\ \Gs_{1,21} &
\Gs_{1,22} \epm\Gc+\bpm \Gs_{2,11} & \Gs_{2,12}\\ \Gs_{2,21} &
\Gs_{2,22} \epm(1-\Gc)\nonum
&=& \bpm \Gs_{1,11} & 0 \\ 0 & 0 \epm\Gc + \bpm 0 & 0 \\ 0 &
\Gs_{1,22} \epm\Gc\nonum
&& +\bpm \Gs_{1,12} & 0 \\ 0 & 0\epm\bpm 0 & 1\\-1 & 0 \epm\Gc -
\bpm 0 & 0 \\ 0 & \Gs_{1,21} \epm\bpm 0 & 1\\ -1 & 0\epm\Gc\\
&& +\bpm\Gs_{2,11} & 0\\ 0 & 0\epm(1-\Gc)+\bpm0 & 0 \\ 0 &
\Gs_{2,22} \epm(1-\Gc)\nonum
&&+\bpm\Gs_{2,12} & 0\\ 0 & 0\epm\bpm0 & 1 \\ -1 & 0\epm(1-\Gc)\nonum
&&-
\bpm 0 & 0\\ 0 & \Gs_{2,21}\epm\bpm0 & 1\\ -1 & 0\epm(1-\Gc)\nonum
&=&
\Gs_{1,11}\BP_1+\Gs_{1,22}\BP_2+\Gs_{1,12}\BP_1\BR_\perp-\Gs_{1,21}\BP_2\BR_\perp\nonum
&&+\Gs_{2,11}\BP_3 +
\Gs_{2,22}\BP_4+\Gs_{2,12}\BP_3\BR_\perp-\Gs_{2,21}\BP_4\BR_\perp.
\eeqa{185}
Let $\BGX$ denote the projection onto $\CU_1\oplus\CE\oplus\CU_2$:
\beq
\BGX = \BGL_1+\bpm0\\\Bu_0\epm\bpm 0 & \Bu_0\epm = \BGL_1 +\bpm 0 &
0\\ 0 & \Bu_0\otimes\Bu_0\epm,
\eeq{186}
in which
\beq
\bpm 0\\\Bu_0\epm = \BR_\perp\bpm\Bu_0\\0\epm
\eeq{187}
lies in the space $\CU_2$. From the equation $\Bj = \BGs\Be$ we have
\beq
\langle \Bj \rangle = \BGX\Bj = \BGX\BGs\Be = \BGX\BGs\BGX\Be\iff
\Be = (\BGX\BGs\BGX)^{-1}\langle\Bj\rangle,
\eeq{188}
where the inverse is to be taken on the space $\CU_1\oplus\CE\oplus\CU_2$.
So if $\BGG_0$ denotes the projection onto $\CU_1\oplus\CU_2$ we have
\beq
\langle \Be \rangle =
\BGG_0(\BGX\BGs\BGX)^{-1}\BGG_0\langle\Bj\rangle =
(\BGs^*)^{-1}\langle\Bj\rangle \iff (\BGs^*)^{-1} =
\BGG_0(\BGX\BGs\BGX)^{-1}\BGG_0.
\eeq{189}
Now we have
\beqa
\BGX\BP_1\BGX &=& \bpm \BI & 0 & 0 & 0\\ 0 & \BI & 0 & 0\\ 0 & 0 &
\BGb\otimes\BGb& 0\\ 0 & 0 & 0& 0\epm\BP_1\bpm \BI & 0 & 0 & 0\\ 0 & \BI & 0 & 0\\ 0 & 0 &
\BGb\otimes\BGb& 0\\ 0 & 0 & 0 & 0\epm\nonum
&=& \bpm \BY_1 & 0 & 0\\
0 & \BY_2 & -\BY_2\BQ^{-1}\BGb\\
0 & -\BGb^T\BQ^{-1}\BY_2 &
\BGb^T\BQ^{-1}\BY_2\BQ^{-1}\BGb\epm\quad\begin{minipage}{0.25\textwidth}in the space \\$\CU_1\oplus\CU_2\oplus\CE,$\end{minipage}
\eeqa{191}
and similarly, in the space $\CU_1\oplus\CE\oplus\CU_2$,
\beq
\BGX\BP_2\BGX = \bpm \BQ^{-1}\BY_2\BQ^{-1} & 0 & 0\\ 0 & \BQ\BY_1\BQ
& -\BQ\BY_1\BGb\\ 0 & -\BGb^T\BY_1\BQ &
\BGb^T\BY_1\BGb\epm.
\eeq{192}
Also since
\beq
\BP_1\BR_\perp = \bpm 0 & \BY_1\BQ & -\BY_1 & 0\\ -\BY_2\BQ^{-1} & 0 & 0
& -\BY_2\\ \BQ^{-1}\BY_2\BQ^{-1} & 0 & 0 & \BQ^{-1}\BY_2\\
0 & \BQ\BY\BQ & -\BQ\BY_1 & 0\epm,
\eeq{193}
and
\beq
\BP_2\BR_\perp = \bpm 0 & \BQ^{-1}\BY_2 & -\BQ^{-1}\BY_2\BQ^{-1} & 0\\
-\BQ\BY_1 & 0 & 0 & -\BQ\BY_1\BQ\\ \BY_1 & 0 & 0 & \BY_1\BQ\\ 0 & \BY_2
& -\BY_2\BQ^{-1} & 0\epm,
\eeq{194}
it follows that in the space $\CU_1\oplus\CE\oplus\CU_2$,
\beqa
\BGX\BP_1\BR_\perp\BGX & = & \bpm 0 & \BY_1\BQ & -\BY_1\BGb\\
                            -\BY_2\BQ^{-1} & 0 & 0\\ 
                             \BGb^T\BQ^{-1}\BY_2\BQ^{-1} & 0 & 0\epm, \nonum
\BGX\BP_2\BR_\perp\BGX & = & \bpm 0 & \BQ^{-1}\BY_2 & -\BQ^{-1}\BY_2\BQ^{-1}\BGb\\ 
                           -\BQ\BY_1 & 0 & 0\\
                            \BGb^T\BY_1 & 0 & 0\epm.
\eeqa{195}
Similarly we have that in the space $\CU_1\oplus\CE\oplus\CU_2$,
\beqa
\BGX\BP_3\BGX &=& \bpm \BY_3 & 0 & 0\\ 
                       0 & \BY_4 & -\BY_4\BQ\BGb\\ 
                       0 & -\BGb^T\BQ\BY_4 & \BGb^T\BQ\BY_4\BQ\BGb\epm,\nonum
\BGX\BP_4\BGX &=& \bpm \BQ\BY_4\BQ & 0 & 0 \\ 
                       0 & \BQ^{-1}\BY_3\BQ^{-1} & -\BQ^{-1}\BY_3\BGb\\ 
                       0 & -\BGb^T\BY_3\BQ^{-1} & \BGb^T\BY_3\BGb\epm,\nonum
\BGX\BP_3\BR_\perp\BGX &=& \bpm       0 & \BY_3\BQ^{-1} & -\BY_3\BGb\\
                              -\BY_4\BQ & 0 & 0\\
                              \BGb^T\BQ\BY_4\BQ & 0 & 0\epm,\nonum
\text{and }\BGX\BP_4\BR_\perp\BGX &=&\bpm           0 & \BQ\BY_4 & -\BQ\BY_4\BQ\BGb\\
                                        -\BQ^{-1}\BY_3 & 0 & 0\\ 
                                    \BGb^T\BY_3 & 0 & 0\epm.
\eeqa{200}
Let $\BA$ denote the matrix
\beqa
\BA &=& \BGX\BGs\BGX\nonum
&=& \Gs_{1,11}\BGX\BP_1\BGX +
\Gs_{1,22}\BGX\BP_2\BGX\nonum
&&+\Gs_{1,12}\BGX\BP_1\BR_\perp\BGX-\Gs_{1,21}\BGX\BP_2\BR_\perp\BGX\nonum
&&+\Gs_{2,11}\BGX\BP_3\BGX+\Gs_{2,22}\BGX\BP_4\BGX\nonum
&&+\Gs_{2,12}\BGX\BP_3\BR_\perp\BGX-\Gs_{2,21}\BGX\BP_4\BR_\perp\BGX.
\eeqa{205}
where the preceeding formulae give the $3\times 3$ block representations in the space 
$\CU_1\oplus\CE\oplus\CU_2$ of the operators needed to compute this matrix.
Let $\BB=\BA^{-1}$ denote the inverse of $\BA$ and let us represent it
in the $3\times 3$ block form
\beq
\BB = \bpm \BB_{11} & \BB_{12} & \Bb_{1}\\
\BB_{21} & \BB_{22} & \Bb_2\\
\Bb_3^T & \Bb_4^T & b\epm,
\eeq{206}
where $\BB_{ij}$ for $i,j=1,2,$ are $(m/2)\times(m/2)$
matrices, $\Bb_i$ for $i=1,2,3,4,$ are $(m/2)-$component vectors,
and $b$ is a scalar. The operator $\BGG_0$ considered as mapping $\CU_1\oplus\CU_2$ to $\CU_1\oplus\CE\oplus\CU_2$
is represented by the matrix
\beq \bpm \BGb & 0\\
0 & 0 \\ 0 & 1\epm, \eeq{206a}
where it takes this form due to the different bases that are used to represent the field $\BU_1$. On the other hand,
when considered as mapping $\CU_1\oplus\CE\oplus\CU_2$ to $\CU_1\oplus\CU_2$, $\BGG_0$ is represented by the matrix
\beq \bpm \BGb^T & 0 & 0\\ 0 & 0 & 1\epm. \eeq{206b}

Thus the effective conductivity tensor function is given by
\beqa
[\BGs^*(\BGs_1,\BGs_2)]^{-1} &=& \bpm \BGb^T & 0 & 0\\ 0 & 0 & 1\epm\BB\bpm \BGb & 0\\
0 & 0 \\ 0 & 1\epm\\
&=&\bpm \BGb\cdot\BB_{11}\BGb & \Bb_1\cdot\BGb\\ \Bb_3\cdot\BGb & b
\epm,
\eeqa{208}
which implicitly gives us the function $\BGs^*(\BGs_1,\BGs_2)$.
\vskip3mm
{\bf Theorem 2}
\vskip2mm
{\it Suppose the conductivity tensor has the form \eq{185}, and consider the domain $\CD(c_1,c_2)$ of pairs $(\BGs_1,\BGs_2)$,
introduced in the appendix, such that \eq{a.000} holds, where $c_1$, $c_2$ are fixed real constants with $c_2>c_1>0$. 
Subject to Assumptions 1, 2, and 3, the effective conductivity tensor function $\BGs_*(\BGs_1,\BGs_2)$
can be approximated arbitrarily closely on $\CD(c_1,c_2)$ by a function that is
expressible in terms of positive definite diagonal $\frac{m}{2}\times\frac{m}{2}$ matrices $\BZ_1$ and $\BZ_2=\BI-\BZ_2$, a $m/2$-component vector $\BGb$ with non-negative
entries, a $n_1\times (\frac{m}{2}-n_1)$ matrix $\BH_1$ satisfying \eq{176} (in which the $\Gr_i$ are the diagonal elements of $\BZ_1$ and the $\Gb_i$ are the components
of the vector $\BGb$), and a $n_2\times (\frac{m}{2}-n_2)$ matrix $\BH_2$ satisfying \eq{178}. The approximating function is given by \eq{208}, with entries obtained through \eq{206a}, where $\BB$ is the inverse of the matrix $\BA$ given in \eq{205},
in which one needs to make the substitutions \eq{191},\eq{192}, \eq{195}, and \eq{200} to express the function in terms of $\BGb$,
$\BQ=\BZ_1^{1/2}\BZ_2^{1/2}$ and $\BY_1$, $\BY_2$, $\BY_3$, and $\BY_4$. The $\BY_i$ are in turn expressible in terms of $\BZ_1$, $\BZ_2$, $\BH_1$, and $\BH_2$
through \eq{159} and \eq{161}. }
\vskip3mm

The challenging question now becomes whether, in the generic case, we
can recover the parameters of the matrices $\BH_1$ and $\BH_2$, in addition to the 
parameters $\Gb_i$ and $\Gr_i$ from knowledge of the function 
$\BGs^*(\BGs_1,\BGs_2)$. Also the question arises as to whether for any $m$, $n_1$ and $n_2$,
with $m/2\geq n_1\geq 1$ and $m/2\geq n_2\geq 1$, for any
set of positive parameters $\Gb_i$ and $\Gr_i$, $i=1,2,\ldots,m/2$, with say
$1> \Gr_i >0$, for any parameters entering the first $n_1-1$ rows of a $n_1\times (\frac{m}{2}-n_1)$ matrix $\BH_1$, and 
for any parameters entering the first $n_2-1$ rows of a $n_2\times (\frac{m}{2}-n_2)$ matrix $\BH_2$ (with the last row in each case being
determined by \eq{176} and \eq{178})
there corresponds, at least in the generic case, a geometry (such as a multiple rank laminate of the two phases) that has
those given parameters and matrices entering the representation formulas of the operators
for that geometry. Alternatively, do there exist some less than obvious restrictions on these
parameters and matrices that hold for any two-phase mixtures?

\section{Approximate formulae for the effective tensors, as functions of the component tensors, for coupled field problems such as elasticity.}
\setcounter{equation}{0}
As mentioned in the introduction, one of the appealing features of our approach is that once one has approximations
for the relevant operators, it allows one to directly obtain approximations for the 
effective tensors $\BL^*$ associated with coupled field problems, with two phases having tensors $\BL_1$ and 
$\BL_2$. Rather considering the general case where the potential $\BV$ has $k$-components and the constitutive equation takes the form \eq{0.0ae} and the fields satisfy \eq{0.0af}, let us confine attention to the case of two-component potentials $\BV=(V_1,V_2)$ (with $k=2$)
as the extension to larger values of $k$ will be clear. The case $k=2$ includes the important cases of two-dimensional elasticity,
thermoelectricity, and magnetoelectricity. (Notice that for  two-dimensional elasticity it is not necessary to introduce the strain field
as the elasticity tensor acting on the displacement gradient $\Grad\BV$ automatically removes the antisymmetric part of $\Grad\BV$ because of the symmetries
of the elasticity tensor).

In the case $k=2$ we introduce the spaces
\beqa
\hat\CP_1 & = & \text{all vector fields of the form }\bpm f_1(\Bx)\\ 0 \\ r_1(\Bx)\\ 0\epm,\nonum
\hat\CP_2 & = & \text{all vector fields of the form }\bpm 0\\ g_1(\Bx) \\ 0\\ s_1(\Bx) \epm,\nonum
\hat\CP_3 & = & \text{all vector fields of the form }\bpm f_2(\Bx)\\ 0 \\ r_2(\Bx)\\0 \epm,\nonum
\hat\CP_4 & = & \text{all vector fields of the form }\bpm 0\\ g_2(\Bx)\\ 0 \\ s_2(\Bx) \epm,\nonum
\eeqa{400}
where $f_1(\Bx)$, $g_1(\Bx)$, $r_1(\Bx)$ and $s_1(\Bx)$ are zero in phase $2$ while 
$f_2(\Bx)$, $g_2(\Bx)$, $r_2(\Bx)$ and $s_2(\Bx)$ are zero in phase $1$. We also let
\beqa
&&\hat\CU_1=\text{the two-dimensional space of fields of the form $\bpm
e_1\\0 \\ e_2 \\ 0 \epm$},\nonum
&&\hat\CU_2=\text{the two-dimensional space of fields of the form $\bpm
0\\e_3\\ 0 \\e_4\epm$},
\eeqa{400aa}
where $e_1,e_2,e_3$ and $e_4$ are constants independent of $\Bx$, and we define
\beqa
\hat\CE  = \left\{\quad\begin{minipage}{0.8\textwidth} 
pairs of curl-free fields which derive from periodic potentials,\\
i.e.\@ fields of the form $\bpm \Grad\Gf_1 \\ \Grad\Gf_2 \epm$ for periodic $\Gf_1(\Bx)$ and $\Gf_2(\Bx)$,
\end{minipage}\right.\nonum
\hat\CJ = \left\{\quad\begin{minipage}{0.8\textwidth}
pairs of divergence-free fields which derive from periodic potentials,\\
i.e.\@ fields of the form $\bpm \BR_\perp\Grad\Gy_1 \\ \BR_\perp\Grad\Gy_2 \epm$
for periodic $\Gy_1(\Bx)$ and $\Gy_2(\Bx)$.
\end{minipage}\right.
\eeqa{400a}
We take as our basis the $4m$ fields
\beq \bpm \Bu_i \\ 0  \epm, \quad \bpm \Bv_i \\ 0  \epm, \quad \bpm 0\\ \Bu_i  \epm, \quad \bpm 0 \\ \Bu_i  \epm, 
\eeq{401} 
with $i$ ranging from $1$ to $m$. They are ordered with the first fields appearing in \eq{401}, with $i=1,2,\ldots, m/2$,
coming first, 
followed by the second fields appearing in \eq{401}, with $i=1,2,\ldots, m/2$,
coming second, then the first fields 
appearing in \eq{401}, with $i=m/2+1,m/2+2,\ldots, m$, come third, and the second fields 
appearing in \eq{401}, with $i=m/2+1,m/2+2,\ldots, m$, come fourth. In fifth position
are the third fields appearing in \eq{401}, with $i=1,2,\ldots, m/2$, followed in sixth position by the
fourth fields appearing in \eq{401}, with $i=1,2,\ldots, m/2$. Finally, in seventh and eighth position
are the third and fourth fields in \eq{401}, respectively, with $i=m/2+1,m/2+2,\ldots, m$.

Accordingly, with respect to this basis the operators $\hat{\BP}_1$, $\hat{\BP}_2$, $\hat{\BP}_3$, $\hat{\BP}_4$
that project onto the four spaces $\hat\CP_1$, $\hat\CP_2$, $\hat\CP_3 $, and $\hat\CP_4$, 
the operators $\hat\BGL_1$ and $\hat\BGL_2$ that project onto the spaces $\hat\CU_1\oplus\hat\CE$
and $\hat\CU_2\oplus\hat\CJ$, and the
operator $\hat\BR_\perp$ that simultaneously rotates the pair of fields by $90^\circ$ are represented by the matrices
\beqa \hat{\BP}_1 & = &\bpm \BP_1 & 0 \\ 0 & \BP_1 \epm, \quad \hat{\BP}_2=\bpm \BP_2 & 0 \\ 0 & \BP_2 \epm, \quad\nonum
\hat{\BP}_3& = &\bpm \BP_3 & 0 \\ 0 & \BP_3 \epm, \quad \hat{\BP}_4=\bpm \BP_4 & 0 \\ 0 & \BP_4 \epm, \quad \nonum
\hat{\BGL}_1 & = &\bpm \BGL_1 & 0 \\ 0 & \BGL_1 \epm, \quad \hat{\BGL}_2=\bpm \BGL_2 & 0 \\ 0 & \BGL_2 \epm, \quad
\hat{\BR}_\perp=\bpm \BR_\perp & 0 \\ 0 & \BR_\perp, \epm
\eeqa{405}
where $\BP_1$, $\BP_2$, $\BP_3$, $\BP_4$ are the matrices \eq{154}, while $\BGL_1$, $\BGL_2$, and $\BR_\perp$ are the
matrices \eq{165}. With these substitutions each of the matrices in \eq{405} has an $8\times 8$ block form, with the submatrices
being $m/2\times m/2$ matrices. The projection onto $\hat\CU_1$ is represented by the matrix
\beq 
\hat\BGG_0^{(1)}=\bpm \underline{\BGb}\otimes \underline{\BGb} & 0 & 0 & 0 & 0 & 0 & 0 & 0 \\
0 & 0 & 0 & 0 & 0 & 0 & 0 & 0 \\
0 & 0 & 0 & 0 & 0 & 0 & 0 & 0 \\
0 & 0 & 0 & 0 & 0 & 0 & 0 & 0  \\
0 & 0 & 0 & 0 &  \underline{\BGb}\otimes \underline{\BGb} & 0 & 0 & 0 \\
0 & 0 & 0 & 0 & 0 & 0 & 0 & 0 \\
0 & 0 & 0 & 0 & 0 & 0 & 0 & 0  \\
0 & 0 & 0 & 0 & 0 & 0 & 0 & 0 \epm,
\eeq{406}
while projection onto $\hat\CU_2$ is represented by the matrix
\beq 
\hat\BGG_0^{(2)}=\bpm 0 & 0 & 0 & 0 & 0 & 0 & 0 & 0 \\
0 & 0 & 0 & 0 & 0 & 0 & 0 & 0 \\
0 & 0 & \underline{\BGb}\otimes \underline{\BGb}& 0 & 0 & 0 & 0 & 0 \\
0 & 0 & 0 & 0 & 0 & 0 & 0 & 0  \\
0 & 0 & 0 & 0 &  0 & 0 & 0 & 0 \\
0 & 0 & 0 & 0 & 0 & 0 & 0 & 0 \\
0 & 0 & 0 & 0 & 0 & 0 & \underline{\BGb}\otimes \underline{\BGb} & 0  \\
0 & 0 & 0 & 0 & 0 & 0 & 0 & 0 \epm.
\eeq{406a}

Similarly to \eq{185}, the operators $\BGs^{(ij)}$ have the representation
\beqa
\BGs^{(ij)}&=&\Gs^{(ij)}_{1,11}\BP_1+\Gs^{(ij)}_{1,22}\BP_2+\Gs^{(ij)}_{1,12}\BP_1\BR_\perp-\Gs^{(ij)}_{1,21}\BP_2\BR_\perp\nonum
&~&+\Gs^{(ij)}_{2,11}\BP_3 +
\Gs^{(ij)}_{2,22}\BP_4+\Gs^{(ij)}_{2,12}\BP_3\BR_\perp-\Gs^{(ij)}_{2,21}\BP_4\BR_\perp.
\eeqa{410}

So analogously to \eq{205} we should introduce the matrix
\beq \BA=\bpm \BA^{(11)} & \BA^{(12)} \\ \BA^{(21)} & \BA^{(22)}\epm,
\eeq{415}
with block matrices
\beqa
\BA^{(ij)}&=& \Gs^{(ij)}_{1,11}\BGX\BP_1\BGX+\Gs^{(ij)}_{1,22}\BGX\BP_2\BGX+\Gs^{(ij)}_{1,12}\BGX\BP_1\BR_\perp\BGX
-\Gs^{(ij)}_{1,21}\BGX\BP_2\BR_\perp\BGX\nonum
&~&+\Gs^{(ij)}_{2,11}\BGX\BP_3\BGX +
\Gs^{(ij)}_{2,22}\BGX\BP_4\BGX+\Gs^{(ij)}_{2,12}\BGX\BP_3\BR_\perp\BGX-\Gs^{(ij)}_{2,21}\BGX\BP_4\BR_\perp\BGX. \nonum
&~&
\eeqa{420}
in which the eight matrices $\BGX\BP_1\BGX$, $\BGX\BP_2\BGX$, $\BGX\BP_1\BR_\perp\BGX$, $\BGX\BP_2\BR_\perp\BGX$,
$\BGX\BP_3\BGX$,  $\BGX\BP_4\BGX$, $\BGX\BP_3\BR_\perp\BGX$, and $\BGX\BP_4\BR_\perp\BGX$ are given by
\eq{191}, \eq{192}, \eq{195}, and \eq{200}. Let $\BB$ denote the inverse of $\BA$ and express it in the block form
\beq
\BB = \bpm \BB_{11}^{(11)}& \BB_{12}^{(11)} & \Bb_{1}^{(11)} & \BB_{11}^{(12)}  & \BB_{12}^{(12)}  & \Bb_{1}^{(12)} \\
\BB_{21}^{(11)}& \BB_{22}^{(11)} & \Bb_2^{(11)} &    \BB_{21}^{(12)}  & \BB_{22}^{(12)}  & \Bb_2^{(12)}  \\
{\Bb_3^{(11)}}^T & {\Bb_4^{(11)}}^T & b^{(11)} & {\Bb_3^{(12)}}^T & {\Bb_4^{(12)}}^T & b^{(12)} \\
\BB_{11}^{(21)} & \BB_{12}^{(21)} & \Bb_{1}^{(21)} & \BB_{11}^{(22)} & \BB_{12}^{(22)} & \Bb_{1}^{(22)} \\
\BB_{21}^{(21)} & \BB_{22}^{(21)} & \Bb_2^{(21)} &    \BB_{21}^{(22)} & \BB_{22}^{(22)} & \Bb_2^{(22)} \\
{\Bb_3^{(21)}}^T & {\Bb_4^{(21)}}^T & b^{(21)}&  {\Bb_3^{(22)}}^T & {\Bb_4^{(22)}}^T & b^{(22)} \epm,
\eeq{421}
where $\BB_{ij}$ for $i,j=1,2,$ are $(m/2)\times(m/2)$
matrices, $\Bb_i$ for $i=1,2,3,4,$ are $(m/2)-$component vectors,
and $b$ is a scalar. Then the effective tensor is given by
\beqa
(\BL^*)^{-1} &=& \bpm \BGb^T & 0 & 0 & 0 & 0 & 0 \\
0 & 0 & 1 & 0 & 0 & 0  \\
0 & 0 & 0 & \BGb^T & 0 & 0 \\
0 & 0 & 0 & 0 & 0 & 1  \\
\epm\BB\bpm \BGb & 0  & 0 & 0\\
0 & 0 & 0 & 0 \\ 0 & 1 & 0 & 0 \\
0 & 0  & \BGb & 0\\
0 & 0 & 0 & 0 \\ 0 & 0 & 0 & 1 
\epm\\
&=&\bpm \BGb\cdot\BB_{11}^{(11)}\BGb & \Bb_1^{(11)}\cdot\BGb & \BGb\cdot\BB_{11}^{(12)}\BGb & \Bb_1^{(12)}\cdot\BGb
\\ \Bb_3^{(11)}\cdot\BGb & b^{(11)} & \Bb_3^{(12)}\cdot\BGb & b^{(12)} \\
\BGb\cdot\BB_{11}^{(21)}\BGb & \Bb_1^{(21)}\cdot\BGb & \BGb\cdot\BB_{11}^{(22)}\BGb & \Bb_1^{(22)}\cdot\BGb\\
\Bb_3^{(21)}\cdot\BGb & b^{(21)} & \Bb_3^{(22)}\cdot\BGb & b^{(22)}
\epm,
\eeqa{430}
which implicitly gives us the function $\BL^*(\BL_1,\BL_2)$.

\section*{Acknowledgements}
The author is grateful to the National Science Foundation for support through the Research Grant DMS-1211359
and to Patrick Bardsley for help in preparing the manuscript. A referee and the associate editor are 
thanked for providing a multitude of helpful comments.

\appendix

\section{Appendix: Truncation of the Hilbert Space}

Consider a domain $\CD(c_1,c_2)$ of $(\BGs_1,\BGs_2)$ pairs, such that for all complex
vectors $\Ba$, with complex conjugate $\overline{\Ba}$,
\beq \Real(\overline{\Ba}\cdot\BGs_i\Ba) \geq c_1|\Ba|^2,\quad |\BGs_i\Ba|\leq c_2|\Ba|,\quad i=1,2, \eeq{a.000}
in which $c_1$ and $c_2$ are fixed constants, with $c_2\geq c_1>0$, $|\Ba|=\sqrt{\overline{\Ba}\cdot\Ba}$, and for any
complex two-component vectors $\Ba=(a_1,a_2)^T$ and $\Bb=(b_1,b_2)^T$ we define $\Ba\cdot\Bb=a_1b_1+a_2b_2$. In the case
where $\BGs_1$ and $\BGs_2$ take the diagonal forms implied by \eq{1}, the restriction \eq{a.000} implies \eq{115aa}.

Here we show how the infinite-dimensional Hilbert space can be truncated to a finite-dimensional one with little change
to the effective conductivity function $\BGs^*(\BGs_1,\BGs_2)$ in the domain $\CD(c_1,c_2)$ of $(\BGs_1,\BGs_2)$ pairs.
The proof is based upon that in Section 3 of \cite{Clark:1994:MEC}. The basic
idea is to show that the Hilbert space can be truncated in such a way that the coefficients in the 
series expansion of $\BGs^*(\BGs_1,\BGs_2)$ about the point $\BGs_1=\BGs_2=\BI$ remain unchanged
up to an arbitrarily large order in the expansion. Rather than doing the truncation in the restricted setting of
Section 3 where $\BGs_1$ and $\BGs_2$ are diagonal, we choose to proceed in the more general setting of section 4, where $\BGs_1$ and $\BGs_2$ need not be diagonal
nor even symmetric. In this setting, the relevant spaces $\CP_i$, $i=1,2,3,4$  are defined by \eq{149a} and the projections $\BP_i$ onto them are defined by \eq{150}. 

Let us relabel the spaces so
\beq \underline{\CH}=\CU\oplus\underline{\CE}\oplus\underline{\CJ}=\underline{\CP}_1\oplus\underline{\CP}_2\oplus\underline{\CP}_3\oplus\underline{\CP}_4,
\eeq{a.0}
is the actual physical infinite-dimensional Hilbert space of interest, where we have introduced underlines on the spaces to distinguish them from
the truncated spaces. We still let $\BP_i$ denote the projection onto $\underline{\CP}_1$, $\BR_\perp$ as that operator which performs a local
rotation of the fields by $90^\circ$, $\BGP$ as that operator associated with reflection,
and $\BGG_0$ as the projection onto $\CU$. However we will label
$\underline{\BGG}_1$ and $\underline{\BGG}_2$ as the projections onto $\underline{\CE}$ and $\underline{\CJ}$ as we will
need slightly different operators $\BGG_1$ and $\BGG_2$ when we define the truncated Hilbert space.

Recalling that $\BU_1\in\CU_1$ and defining $\BU_2=\BR_\perp\BU_1\in\CU_2$, let us introduce the multi-index fields
\beqa \BE_{\Ga_m j} & = &\underline{\BGG}_1\BP_{a_1}\underline{\BGG}_1\BP_{a_2}\underline{\BGG}_1\BP_{a_3}\ldots\underline{\BGG}_1\BP_{a_m}\BU_j, \nonum
 \BJ_{\Ga_m j} & = & \underline{\BGG}_2\BP_{a_1}\underline{\BGG}_2\BP_{a_2}\underline{\BGG}_2\BP_{a_3}\ldots\underline{\BGG}_2\BP_{a_m}\BU_j,
\eeqa{a.1}
where $\Ga_m=(a_1,a_2,a_3,\ldots,a_m)$ is a multi-index comprised of indices $a_1,a_2,a_3,\ldots,a_m$ in which $m$ will be called
the order of $\Ga_m$. Thus for a given order $m$, $\Ga_m$ can take $4^m$ different values and we denote the set of these values as $\CA_m$. 
We define the subspaces

\vskip4mm

$\bullet$ $\tilde{\CE}=$ the space spanned by the fields $\BE_{\Ga_m j}$, $j=1,2$, as $a_m$ ranges over all combinations in $\CA_m$ and
$m$ ranges from $1$ to some maximum value $m=M$. 

\vskip4mm

$\bullet$ $\tilde{\CJ}=$ the space spanned by the fields $\BJ_{\Ga_m j}$, $j=1,2$, as $a_m$ ranges over all combinations in $\CA_m$ and
$m$ ranges from $1$ to some maximum value $m=M$.

\vskip4mm

The space $\tilde{\CH}\equiv \CU\oplus\tilde{\CE}\oplus\tilde{\CJ}$ is closed under the action of $\underline{\BGG}_1$, $\underline{\BGG}_2$, and $\BR_\perp$ but not under the action of
the $\BP_i$, $i=1,2,3,4$. To see this, notice that
\beqa \BP_i\BE_{\Ga_m j}&=&(\BGG_0+\underline{\BGG}_1+\underline{\BGG}_2)\BP_i\BE_{\Ga_m j} \nonum
&=&
\BGG_0\BP_i\BE_{\Ga_m j}+\BE_{\Gb_{m+1} j} \nonum 
&~& +
\underline{\BGG}_2\BP_i(\BI-\BGG_0-\underline{\BGG}_2)\BP_{a_1}(\BI-\BGG_0-\underline{\BGG}_2)\BP_{a_2}(\BI-\BGG_0-\underline{\BGG}_2)\BP_{a_3}
\ldots(\BI-\BGG_0-\underline{\BGG}_2)\BP_{a_m}\BU_j, \nonum
\eeqa{a.1a}
where $\Gb_{m+1}$ is the multi-index $\Gb_{m+1}=(i,a_1,a_2,a_3,\ldots,a_m)$, lies in $\tilde{\CH}$ for $m<M$ but not for $m=M$. Similarly,
$\BP_i\BJ_{\Ga_m j}$ lies in $\tilde{\CH}$ for $m<M$ but not for $m=M$. The fields in $\tilde{\CH}$ are precisely those that appear in
the series expansions up to order $M$ for the fields $\Be(\Bx)$ and $\Bj(\Bx)$ that solve \eq{0.0aa} and \eq{0.0} when $\BGs_1$ and $\BGs_2$ are both
close to the identity matrix $\BI$, and this motivates their introduction. Note we do not assume that the set of fields $\BE_{\Ga_m j}$ (nor $\BJ_{\Ga_m j}$)
are independent, i.e., some could be linear combinations of the others.

Let $\BGY$ denote the projection onto $\tilde{\CH}$ and define the fields
\beq \Br_{i\Ga_M}=(\BI-\BGY)\BP_i\BE_{\Ga_M 1}, \quad \Br^\perp_{i\Ga_M}=\BR_{\perp}\Br_{i\Ga_M}.
\eeq{a.2}
Let ${\CR}$ and ${\CR}_\perp$ be the spaces spanned by $\Br_{i\Ga_M}$ and $\Br^\perp_{i\Ga_M}$, respectively, 
as $i$ and $\Ga_M$ vary in the sets $\{1,2,3,4\}$ and $\CA_M$. The space
$\CH=\CU\oplus\tilde{\CE}\oplus\tilde{\CJ}\oplus{\CR}\oplus{\CR}_\perp$ is closed under the action of $\BR_\perp$, $\BGP$, and $\BP_i$, $i=1,2,3,4$.
Closure under $\BR_\perp$ is clear since
\beq \BR_\perp\CU=\CU,\quad \BR_\perp\tilde{\CE}=\tilde{\CJ}, \quad\BR_\perp\tilde{\CJ}=\tilde{\CE}, \quad\BR_\perp{\CR}={\CR}_\perp,
\quad \BR_\perp{\CR}_\perp={\CR},
\eeq{a.3}
and it is similarly easy to see the closure under $\BGP$.
Now for $i=1,2,3,4$ define $i^c$ to be $2,1,4,3$ respectively, and given a multi-index $\Ga_m=(a_1,a_2,a_3,\ldots,a_m)$, define $\Ga_m^c=(a_1^c,a_2^c,a_3^c,\ldots,a_m^c)$.
Then closure of $\CH$ under $\BP_i$ follows from \eq{a.1a} and because
\beqa \BP_i\BE_{\Ga_M 1} & = & \Br_{i\Ga_M}+\BGY\BP_i\BE_{\Ga_M 1}\in\CH, \nonum
\BP_i\BE_{\Ga_M 2} & = & \BP_i\underline{\BGG}_1\BP_{a_1}\underline{\BGG}_1\BP_{a_2}\underline{\BGG}_1\BP_{a_3}\ldots\underline{\BGG}_1\BP_{a_M}\BR_\perp\BU_1 \nonum
                & = & \BR_\perp\BP_{i^c}\underline{\BGG}_2\BP_{a_1^c}\underline{\BGG}_2\BP_{a_2^c}\underline{\BGG}_2\BP_{a_3^c}\ldots\underline{\BGG}_2\BP_{a_M^c}\BU_1 \nonum
 & = & \BR_\perp\BP_{i^c}(\BI-\BGG_0-\underline{\BGG}_1)\BP_{a_1^c}(\BI-\BGG_0-\underline{\BGG}_1)\BP_{a_2^c}(\BI-\BGG_0-\underline{\BGG}_1)\BP_{a_3^c}\ldots(\BI-\BGG_0-\underline{\BGG}_1)\BP_{a_M^c}\BU_1 \nonum
 & = & (-1)^M\BR_\perp\BP_{i^c}\BE_{\Ga_M^c 1}+\text{ fields in }\tilde{\CH}\in\CH, \nonum
\BP_i\BJ_{\Ga_M 1} & = & \BP_i\underline{\BGG}_2\BP_{a_1}\underline{\BGG}_2\BP_{a_2}\underline{\BGG}_2\BP_{a_3}\ldots\underline{\BGG}_2\BP_{a_M}\BU_1 \nonum
                 & = & \BP_i(\BI-\BGG_0-\underline{\BGG}_1)\BP_{a_1}(\BI-\BGG_0-\underline{\BGG}_1)\BP_{a_2}(\BI-\BGG_0-\underline{\BGG}_1)\BP_{a_3}\ldots(\BI-\BGG_0-\underline{\BGG}_1)\BP_{a_M}\BU_1 \nonum
                 &= & (-1)^M\BP_i\BE_{\Ga_M 1} +\text{ fields in }\tilde{\CH}\in\CH, \nonum
\BP_i\BJ_{\Ga_M 2} & = &  \BP_i\underline{\BGG}_2\BP_{a_1}\underline{\BGG}_2\BP_{a_2}\underline{\BGG}_2\BP_{a_3}\ldots\underline{\BGG}_2\BP_{a_M}\BR_\perp\BU_1 \nonum
                & = & \BR_\perp\BP_{i^c}\underline{\BGG}_1\BP_{a_1^c}\underline{\BGG}_1\BP_{a_2^c}\underline{\BGG}_1\BP_{a_3^c}\ldots\underline{\BGG}_1\BP_{a_M^c}\BU_1 \nonum
                & = & \BR_\perp\BP_{i^c}\BE_{\Ga_M^c 1}=\Br^\perp_{i^c\Ga_M^c}+\BGY\BR_\perp\BP_{i^c}\BE_{\Ga_M^c 1} \in \CH.
\eeqa{a.4}
Now we let
\beq \CU=\overline{\CU}, \eeq{a.5}
\beq \CE= \text{ the space spanned by }\tilde{\CE}\text{ and }{\CR}, \eeq{a.6}
\beq \CJ= \text{ the space spanned by }\tilde{\CJ}\text{ and }{\CR}_\perp, \eeq{a.7}
\beq \CP_i=\underline{\CP}_i\cap\CH, \quad i=1,2,3,4. \eeq{a.8}
It is clear that $\CU$, $\CE$, and $\CJ$ are mutually orthogonal and 
\beq \CH=\CU\oplus\CE\oplus\CJ=\CP_1\oplus\CP_2\oplus\CP_3\oplus\CP_4
\eeq{a.9}
defines our truncated Hilbert space. The operators $\BP_i$, $i=1,2,3,4$, remain
unchanged and act within $\CH$ to project on $\CP_i$, $i=1,2,3,4$, respectively.
The projections $\BGG_1$ and $\BGG_2$ that project onto $\CE$ and $\CJ$, respectively,
differ slightly from $\underline{\BGG}_1$ and $\underline{\BGG}_2$ (and do not act locally in Fourier space).

The subsequent proof that this truncation does not effect the series expansion coefficients up to order $M$,
and as a consequence produces very little change to the effective conductivity function $\BGs^*(\BGs_1,\BGs_2)$
in the domain $\CD(c_1,c_2)$ defined by \eq{a.000}, when 
$M$ is large, proceeds essentially the same as detailed in Section 3 of \cite{Clark:1994:MEC}. We will
not repeat this proof here. Of course if $c_1$ is decreased (but still remaining positive) and $c_2$ is increased
then $M$ must be correspondingly increased to get a good approximation in the larger domain $\CD(c_1,c_2)$.

\end{document}